%% file: main.tex
\documentclass[twocolumn]{aastex7}
\usepackage{amsmath}
\usepackage[T1]{fontenc}
\usepackage{upgreek}
\usepackage{comment}

\input{affil_list}

\submitjournal{AJ}

\shorttitle{NETS Three Year Status}
\shortauthors{Gupta et al.}

\graphicspath{{./}{figures/}}

\newcommand{\rev}{}

\begin{document}

\title{The NEID Earth Twin Survey. III. Survey Performance After Three Years on Sky}

\author[0000-0002-5463-9980]{Arvind F.\ Gupta}
\affil{\NOIRLab}
\email{arvind.gupta@noirlab.edu}

\author[0000-0003-0199-9699]{Evan Fitzmaurice}
\affil{\PSUAA}
\affil{\PSUCEHW}
\affil{\PSUICDS}
\email{exf5296@psu.edu}

\author[0000-0001-9596-7983]{Suvrath Mahadevan}
\affiliation{\PSUAA}
\affiliation{\PSUCEHW}
\affiliation{\PSUARC}   
\email{suvrath@astro.psu.edu}

\author[0000-0003-0149-9678]{Paul Robertson}
\affiliation{\UCI}
\email{paul.robertson@uci.edu}

\author[0000-0002-4927-9925]{Jacob K. Luhn}
\affiliation{\JPL}
\email{jacob.luhn@jpl.nasa.gov}

\author[0000-0001-6160-5888]{Jason T.\ Wright}
\affiliation{\PSUAA}
\affiliation{\PSUCEHW}
\affiliation{\PSETI}
\email{astrowright@gmail.com}

\author[0000-0002-9632-9382]{Sarah E.\ Logsdon}
\affiliation{\NOIRLab}
\email{sarah.logsdon@noirlab.edu}

\author[0000-0001-9626-0613]{Daniel M.\ Krolikowski}
\affiliation{\UA}
\email{krolikowski@arizona.edu}

\author[0000-0003-1324-0495]{Leonardo A.\ Paredes}
\affiliation{\UA}
\email{}

\author[0000-0003-4384-7220]{Chad F.\ Bender}
\affiliation{\UA}
\email{cbender@arizona.edu}

\author[0000-0002-0078-5288]{Mark R.~Giovinazzi}
\affiliation{\Amherst}
\email{mgiovinazzi@amherst.edu}

\author[0000-0002-9082-6337]{Andrea S.J.\ Lin}
\affiliation{\Caltech}
\email{asjlin@caltech.edu}


\author[0000-0002-6096-1749]{Cullen H.\ Blake}
\affiliation{\Penn}
\email{chblake@sas.upenn.edu}

\author[0000-0003-4835-0619]{Caleb I. Ca\~nas}
\affiliation{\GoddardESAL}
\email{c.canas@nasa.gov}


\author[0000-0001-6545-639X]{Eric B.\ Ford}
\affiliation{\PSUAA}
\affiliation{\PSUCEHW}
\affiliation{\PSUICDS}
\affiliation{\PSUCASt}
\email{ebf11@psu.edu}

\author[0000-0003-1312-9391]{Samuel Halverson}
\affiliation{\JPL}
\email{samuel.halverson@jpl.nasa.gov}


\author[0000-0001-8401-4300]{Shubham Kanodia}
\affiliation{\Carnegie}
\email{skanodia@carnegiescience.edu}

\author[0000-0003-0241-8956]{Michael W.\ McElwain}
\affiliation{\GoddardESAL} 
\email{michael.w.mcelwain@nasa.gov}

\author[0000-0002-0048-2586]{Andrew Monson}
\affiliation{\UA}
\email{}

\author[0000-0001-8720-5612]{Joe P.\ Ninan}
\affiliation{\TIFR}
\email{indiajoe@gmail.com}

\author[0000-0002-2488-7123]{Jayadev Rajagopal}
\affiliation{\NOIRLab}
\email{}

\author[0000-0001-8127-5775]{Arpita Roy}
\affiliation{Astrophysics \& Space Institute, Schmidt Sciences, New York, NY 10011, USA}
\email{arpita308@gmail.com}

\author[0000-0002-4046-987X]{Christian Schwab}
\affiliation{\Macquarie}
\email{mail.chris.schwab@gmail.com}

\author[0000-0001-7409-5688]{Gudmundur Stefansson}
\affiliation{\UAm}
\email{g.k.stefansson@uva.nl}

\author[0000-0002-4788-8858]{Ryan C. Terrien}
\affiliation{\Carleton}
\email{rterrien@carleton.edu}

\correspondingauthor{Arvind F.\ Gupta}
\email{arvind.gupta@noirlab.edu}

\begin{abstract}
    The NEID Earth Twin Survey (NETS) has been delivering a rich set of precise radial velocity (RV) measurements for 41 bright, nearby main sequence stars. Here, we describe the status of the survey after three years on sky and we present the full set of RV measurements and accompanying stellar activity indicators. We discuss intermediate survey diagnostics, including calibration of the known RV zero point offset introduced following the Contreras fire in 2022 and the identification of an undiagnosed and previously unknown zero point offset in 2021.
    An analysis of our data set using \texttt{RVSearch} demonstrates that for these target stars, NEID is independently sensitive to nearly all known planets with periods shorter than the NETS observing baseline. We also highlight a number of newly detected RV signals, which present exciting opportunities for future investigations.
\end{abstract}


\section{Introduction}\label{sec:intro}

The NEID Earth Twin Survey \citep[NETS;][]{Gupta2021} is a dedicated search for low-mass, long-period exoplanets with NEID \citep{Schwab2016}, an extreme precision radial velocity (EPRV) spectrograph on the WIYN 3.5\,m telescope at Kitt Peak National Observatory.\footnote{The WIYN Observatory is a joint facility of the NSF’s National Optical-Infrared Astronomy Research Laboratory, Indiana University, the University of Wisconsin-Madison, Pennsylvania State University, and Princeton University.} 
Using time allocated through a guaranteed time observations (GTO) program, supplemented in part by Penn State institutional access to NEID, we have been monitoring a carefully selected sample of 41 bright, RV-quiet stars since the commencement of this survey in September 2021. Data from the first few years of the survey have led to the confirmation of a low-mass planet \citep{Gupta2025}, the precise characterization of long-period stellar and planetary orbits \citep{Giovinazzi2025}, and the refinement of stellar activity mitigation methods \citep{Burrows2024,Siegel2024,Gilbertson2024}.

In this work, we describe the current status of NETS observations and the performance of the survey after its first three years. In Section \ref{sec:observations}, we outline the intended NETS observing strategy and cadence and compare these to the acquired observations for each target star. We also discuss emergent issues with the survey and spectrograph and the solutions that have been implemented to address these. We calculate summary statistics for each star in Section \ref{sec:diagnostics} based on curated data from the first three years of observations, including a first order calibration of the RV zero point offset introduced by the instrument shutdown and subsequent recovery in 2022. In Section \ref{sec:rvsearch}, we describe our approach to testing the sensitivity of NETS data to known astrophysical signals in our sample, and we assess our ability to independently recover these in Section \ref{sec:known_signals}. We wrap up in Section \ref{sec:discussion} with a summary of the overall survey sensitivity and a discussion of synergy between NETS and future missions to characterize Earth-like exoplanets.

\section{Observations}\label{sec:observations}
\subsection{NETS Observing Strategy}\label{sec:obs_strategy}

NEID is a fiber-fed \citep{Kanodia2023}, environmentally-stabilized \citep{Robertson2019} optical to near-infrared spectrograph. The instrument was designed to exceed an internal precision requirement  of 27 cm~s$^{-1}$ based on a comprehensive, bottom-up error budget presented by \citet{Halverson2016}. Tests performed during instrument commissioning and data from the NEID solar feed \citep{Lin2022,Ford2024} demonstrate an achieved on-sky precision of $<50$ cm~s$^{-1}$. In addition, thanks to its broad wavelength coverage of more than 380 -- 930 nm, NEID observations cover a number of spectral features that are known to trace stellar magnetic activity, including the Ca II H\&K lines, H$\alpha$, and the Ca II infrared triplet.

NETS observations are designed to leverage the precision capabilities of NEID to enable the detection of exoplanets with sub-m~s$^{-1}$ RV semi-amplitudes. We maintain a relatively uniform observing strategy across the target sample, which we list in \autoref{tab:nets_stats}. That is, most stars are observed to the same single-measurement precision (Section \ref{sec:single_night}),  we schedule the same number of annual observations for every star, and these observations are distributed across each star's observing season in a consistent manner (Section \ref{sec:seasonal}). These choices are made so that the survey will be amenable to demographics studies, unlike RV surveys that do not follow a similar systematic strategy.

\begin{deluxetable*}{lrrrrrrrrr}
\tablecaption{NEID Earth Twin Survey Observing Statistics \label{tab:nets_stats}}
\tablehead{
    \colhead{Star Name}& $\tau$ [d]& \multicolumn{3}{c}{$N_{\rm obs}$} & \multicolumn{3}{c}{RV RMS [m s$^{-1}$]} & \multicolumn{2}{c}{$\Delta\gamma$ [m s$^{-1}$]}\\
    \colhead{} & \colhead{} & \colhead{Run 0.5} & \colhead{Run 1} & \colhead{Run 2} & \colhead{Run 0.5} & \colhead{Run 1} & \colhead{Run 2} & Run 1$-$0.5 & Run 1$-$2
}
\startdata
HD 4614 A & 1070 & 1 & 30 & 58 & \nodata & 1.16 & 1.19 & \nodata & \nodata \\
HD 4628 & 1021 & \nodata & 35 & 69 & \nodata & 1.21 & 1.36 & \nodata & $3.28$  \\
HD 9407 & 1247 & 2 & 39 & 61 & \nodata & 1.16 & 1.10 & $0.71$ & $1.01$ \\
HD 10476 & 878 & \nodata & 26 & 44 & \nodata & 1.21 & 2.03 & \nodata & $0.11$  \\
HD 10700 & 1110 & 1 & 30 & 54 & \nodata & 0.86 & 1.09 & $-0.02$ & $1.68$ \\
HD 10780 & 1002 & \nodata & 32 & 59 & \nodata & 2.12 & 2.51 & \nodata & $-0.18$  \\
HD 19373 & 904 & \nodata & 36 & 49 & \nodata & 1.72 & 1.73 & \nodata & $0.77$  \\
HD 24496 A & 902 & \nodata & 32 & 50 & \nodata & 1.19 & 3.15 & \nodata & \nodata  \\
HD 26965 & 861 & \nodata & 36 & 47 & \nodata & 2.57 & 2.34 & \nodata & $3.70$  \\
HD 34411 & 917& \nodata & 29 & 58 & \nodata & 1.08 & 1.17 & \nodata & $1.66$  \\
HD 38858 & 908& \nodata & 29 & 50 & \nodata & 2.42 & 2.06 & \nodata & $5.22$  \\
HD 50692 & 952& \nodata & 31 & 57 & \nodata & 1.49 & 2.32 & \nodata & $2.84$  \\
HD 51419 & 950& \nodata & 32 & 55 & \nodata & 1.00 & 1.12 & \nodata & $1.63$  \\
HD 52711 & 947& \nodata & 30 & 53 & \nodata & 1.18 & 1.65 & \nodata & $2.18$  \\
HD 55575 & 921& \nodata & 32 & 60 & \nodata & 1.29 & 1.50 & \nodata & $1.28$  \\
HD 68017 & 974& \nodata & 30 & 82 & \nodata & 0.90 & 1.56 & \nodata & \nodata  \\
HD 86728 & 1203& 5 & 38 & 94 & 0.76 & 0.88 & 1.18 & $1.15$ & $1.37$ \\
HD 95735 & 1246& 4 & 34 & 94 & 1.01 & 0.79 & 1.35 & \nodata & \nodata \\
HD 110897 & 970& \nodata & 31 & 83 & \nodata & 1.48 & 2.03 & \nodata & $0.47$  \\
HD 115617 & 1229 & 3 & 23 & 73 & 0.77 & 1.85 & 2.23 & $3.71$ & $2.02$ \\
HD 116442 & 914& \nodata & 22 & 66 & \nodata & 0.74 & 0.61 & \nodata & $0.73$  \\
HD 126053 & 938& \nodata & 25 & 75 & \nodata & 1.46 & 1.41 & \nodata & $1.16$  \\
HD 127334 &1270 & 25 & 44 & 115 & 0.94 & 1.14 & 0.99 & $1.02$ & $1.24$ \\
HD 143761 & 1253& 10 & 27 & 71 & 3.40 & 4.14 & 4.01 & $-1.38$ & $0.50$ \\
HD 146233 & 885 & \nodata & 16 & 63 & \nodata & 1.96 & 2.15 & \nodata & $4.82$  \\
HD 154345 & 1001& \nodata & 23 & 56 & \nodata & 2.38 & 2.97 & \nodata & \nodata  \\
HD 157214 & 1068 & 1 & 23 & 62 & \nodata & 1.07 & 1.06 & $0.35$ & $1.38$ \\
HD 166620 & 1217& 12 & 22 & 58 & 0.62 & 0.82 & 0.90 & $0.78$ & $1.07$ \\
HD 168009 & 1014& \nodata & 31 & 59 & \nodata & 2.81 & 3.16 & \nodata & $2.98$  \\
HD 170657 & 1134& 5 & 10 & 45 & 0.92 & 2.77 & 3.20 & $3.21$ & $1.84$ \\
HD 172051 & 1022& \nodata & 9 & 45 & \nodata & 1.24 & 1.03 & \nodata & $1.25$  \\
HD 179957 & 1001& \nodata & 14 & 28 & \nodata & 0.73 & 1.23 & \nodata & \nodata  \\
HD 182572 & 1012& \nodata & 16 & 59 & \nodata & 2.10 & 1.98 & \nodata & $2.61$  \\
HD 185144 & 1230& 19 & 49 & 95 & 0.73 & 0.92 & 0.60 & $1.82$ & $1.66$ \\
HD 186408 & 944& \nodata & 13 & 41 & \nodata & 2.58 & 1.72 & \nodata & $-0.33$  \\
HD 186427 & 949& \nodata & 13 & 42 & \nodata & 1.68 & 2.30 & \nodata & $0.57$  \\
HD 187923 & 934& \nodata & 12 & 42 & \nodata & 1.88 & 1.54 & \nodata & $1.05$  \\
HD 190360 & 1074 & 5 & 24 & 45 & 1.37 & 1.89 & 2.03 & \nodata & \nodata \\
HD 201091 & 1107& 7 & 30 & 49 & 1.94 & 2.18 & 1.75 & \nodata & \nodata \\
HD 217107 & 1032& \nodata & 22 & 33 & \nodata & 4.66 & 4.54 & \nodata & \nodata  \\
HD 221354 & 1099& 2 & 41 & 65 & \nodata & 1.06 & 1.19 & $0.05$ & $1.36$ \\
\enddata
\tablenotetext{}{\textbf{Note.} Columns are total NEID baseline ($\tau$), number of observations ($N_{\rm obs}$), root mean square RV (RMS), and zero point offset across NEID RV eras ($\Delta\gamma$). Separate values are reported for each run or pair of runs. RMS values are not reported for stars with fewer than three visits in a given run. Zero point offsets are not reported for stars with known RV signals induced by long-period stellar or planetary companions, as described in Section \ref{sec:zeropoint}.}
\end{deluxetable*}

\subsubsection{Nightly Observing Strategy}\label{sec:single_night}

The target single-measurement precision for NETS observations follows that described by \citet{Gupta2021}, which we summarize here. We first calculate how long it will take to reach a combined $\sigma_{\rm{RV},\star} = \sqrt{\sigma_{\rm phot}^2 +\sigma_{\rm osc}^2} \leq 30$ cm s$^{-1}$. The photon noise precision, $\sigma_{\rm phot}$, is estimated for median observing conditions using the NEID exposure time calculator, which interpolates over a grid of exposure time, stellar effective temperature ($T_{\rm eff}$), and apparent $V$-mag
generated following \citet{Bouchy2001}. For the photon noise calculation, the projected rotational velocity, $v \sin i$, was held fixed at 2 km~s$^{-1}$. The residual p-mode oscillation amplitude, $\sigma_{\rm osc}$, is estimated using the model outlined by \citet{Chaplin2019}. We then compare this to the period corresponding to the peak of the oscillation power envelope, $1/\nu_{\rm max}$, and take the larger of these two values to be the nominal exposure time for each star. To avoid incurring additional uncertainty due to variations in the charge transfer inefficiency (CTI) of the NEID detector \citep{Blake2017}, individual exposures are set to trigger at a fixed signal-to-noise ratio (S/N) corresponding to this nominal exposure time. For the brightest stars in our sample, for which the calculated S/N exceeds the detector non-linearity threshold at any wavelength (60\% well depth; S/N$\sim480$ per 1-D extracted pixel), observations are split into multiple consecutive exposures with a total baseline (including inter-exposure readout) equal to the nominal exposure time. We also note that for all NETS observations, we simultaneously feed light from the etalon \citep{Kreider2025} through the calibration fiber, except when the etalon is offline for maintenance. These measurements can be used to improve sampling of the nightly instrument drift. 

The 30 cm s$^{-1}$ precision threshold was selected by \citet{Gupta2021} because it is on par with the internal NEID precision requirement and it can be achieved with reasonably short exposures ($\lesssim 10 $ minutes) for most NETS targets. For a subset of these stars, however, we expect to reach 30 cm s$^{-1}$ with exposure sequences shorter than the $\sim5$-minute overheads associated with each NEID observation. To improve the duty cycle and increase the precision of observations for some of the brightest stars, we double the number of exposures per visit for the stars HD\,4614A, HD\,10476, HD\,26965, HD\,115617, HD\,185144, and HD\,201091, and we triple the number of exposures per visit for HD\,10700. In each of these cases, the total nominal visit duration is still less than 10 minutes.

While NEID commissioning was ongoing and prior to initiating observations for the full NETS sample in September 2021, we observed a limited set of stars from January 2021 through August 2021 to explore the precision capabilities of the spectrograph and the practicality of our intended nightly observing strategy within the constraints of the NEID Queue.\footnote{For details about the NEID Queue, see \url{https://wiyn-queuemaster.kpno.noirlab.edu/manual/}} For these precursor observations, we aimed to schedule two visits per night, separated by a minimum of two hours. This sampling was intended to mitigate the impact of correlated noise contributions from granulation and supergranulation \citep{Dumusque2011}, as these processes have characteristic timescales of hours to days. While these observations were executed as intended in most cases, we were sometimes left with unpaired visits as a result of changing observing conditions and unforeseen technical issues. \rev{In addition, while the measured change in RV within a night was significant for some stars \citep[e.g., HD\,86728, as shown in][]{Gupta2025}, paired measurements were typically consistent to within the photon noise uncertainties for most targets. With fewer than ten paired measurements per star during this pre-survey phase, the results of the test were not conclusive.} In the interest of maintaining a more consistent observing strategy across all nights, we elected to proceed with just a single visit per night once the full survey began. 

\subsubsection{Seasonal Observing Strategy}\label{sec:seasonal}

Given the size of our target sample and the typical cost of a single visit ($\sim15$ min), the annual NETS time allocation was expected to accommodate 36 visits per star per year. We distribute these across the NEID Queue priority levels\footnote{In general, NEID queue time allocations are divided into five priority levels (P0-P4). For more details on NEID's priority-level breakdown see \url{https://www.wiyn.org/Instruments/NEID_PriorityLevelDesignationv2.pdf}} to match the 3:2:1 ratio of time available to the survey at levels 2, 1, and 0, respectively. Each star is therefore allocated 18 Priority 2 (P2) visits, 12 Priority 1 (P1) visits, and 6 Priority 0 (P0) visits, where P0 observations have the highest priority to be scheduled and executed, P1 observations have the next highest priority, and so on through P4.
Penn State institutional access accounts for 21.9\% of the total P0 time, 14.7\% of the total P1 time, and 16.1\% of the total P2 time, with the remainder coming from the GTO program.
To increase the sensitivity of our survey to low-mass, long-period exoplanets, we use the observations in each of these priority bins to build a seasonal observing schedule for each star that mixes high cadence, low cadence, and time-constrained observations.

\begin{figure*} 
    \includegraphics[ width=0.91\linewidth, trim={2cm 0cm 0.3cm 0cm},clip]{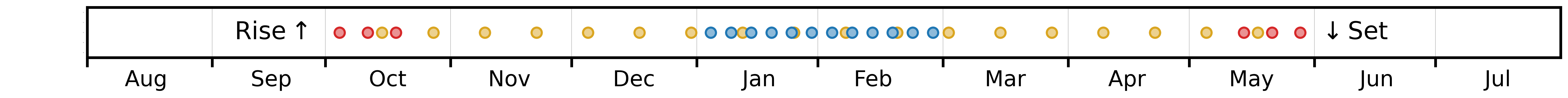}
    \centering
    \includegraphics[width=0.95\linewidth]{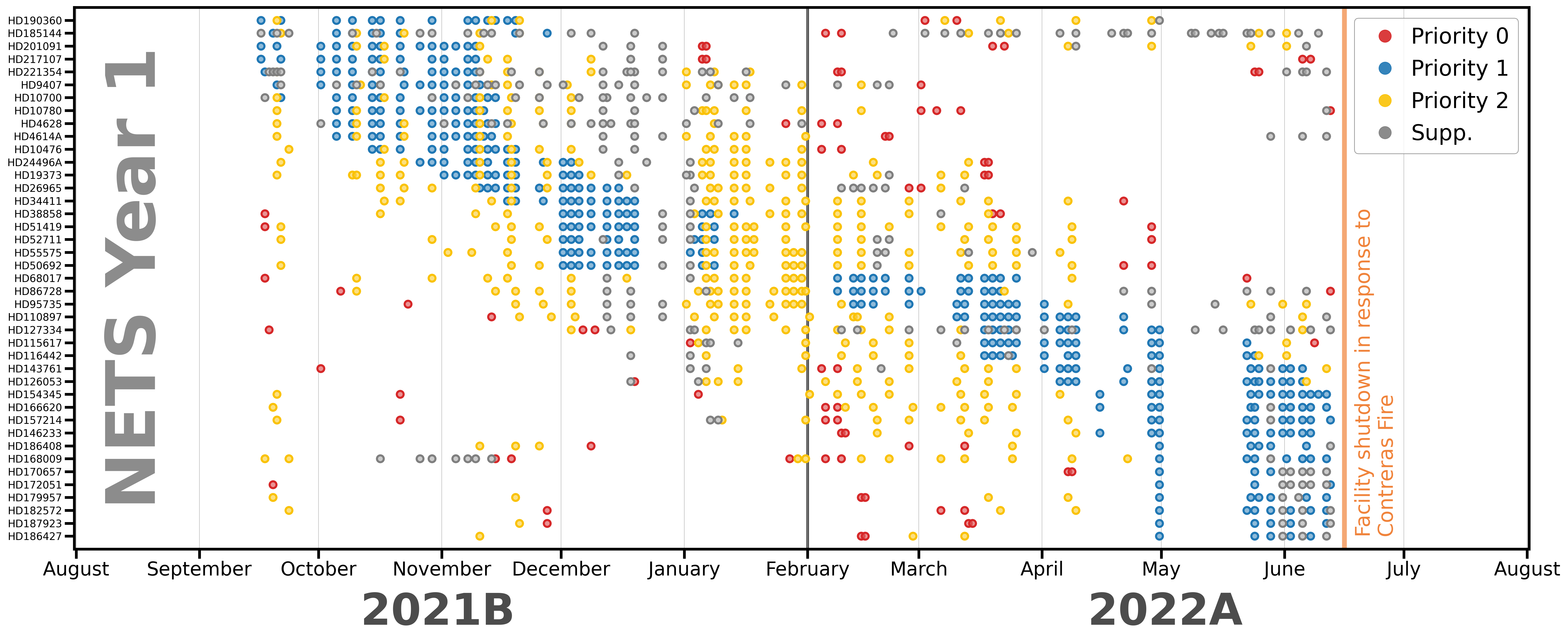}
    \includegraphics[width=0.95\linewidth]{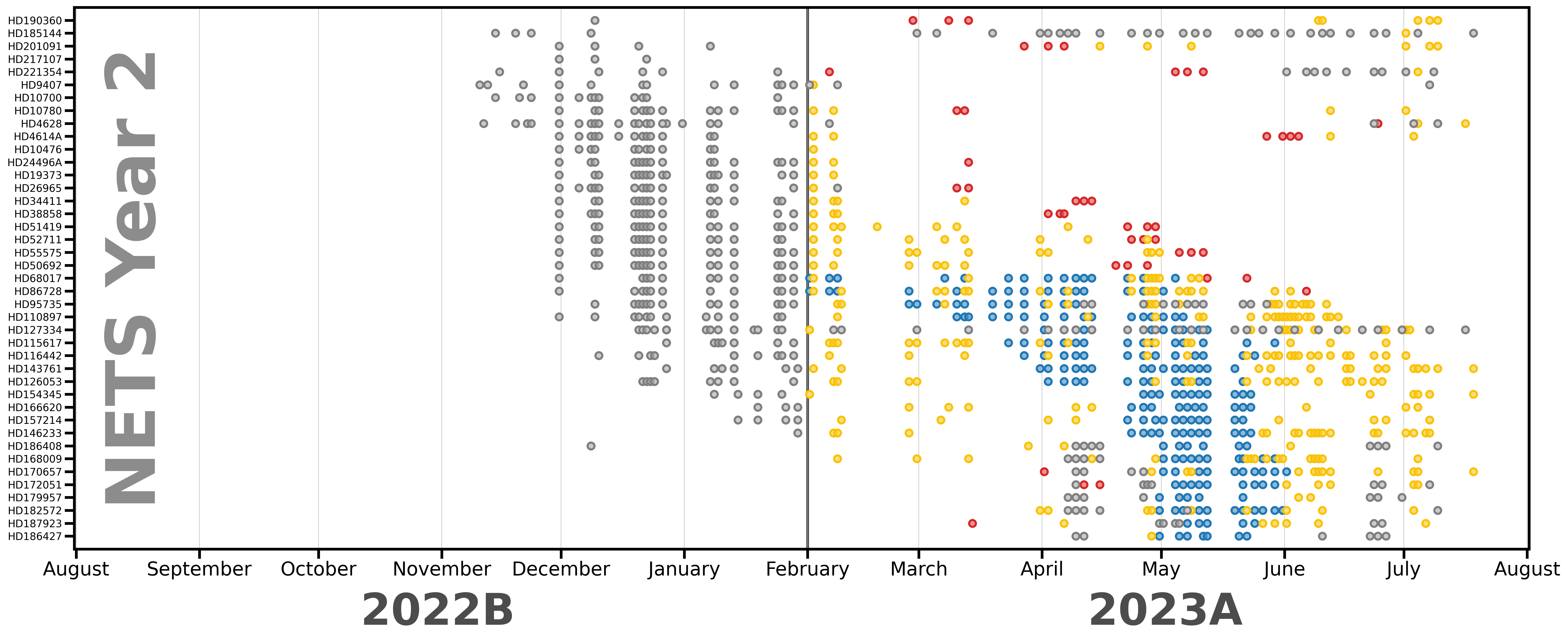}
    \includegraphics[width=0.95\linewidth]{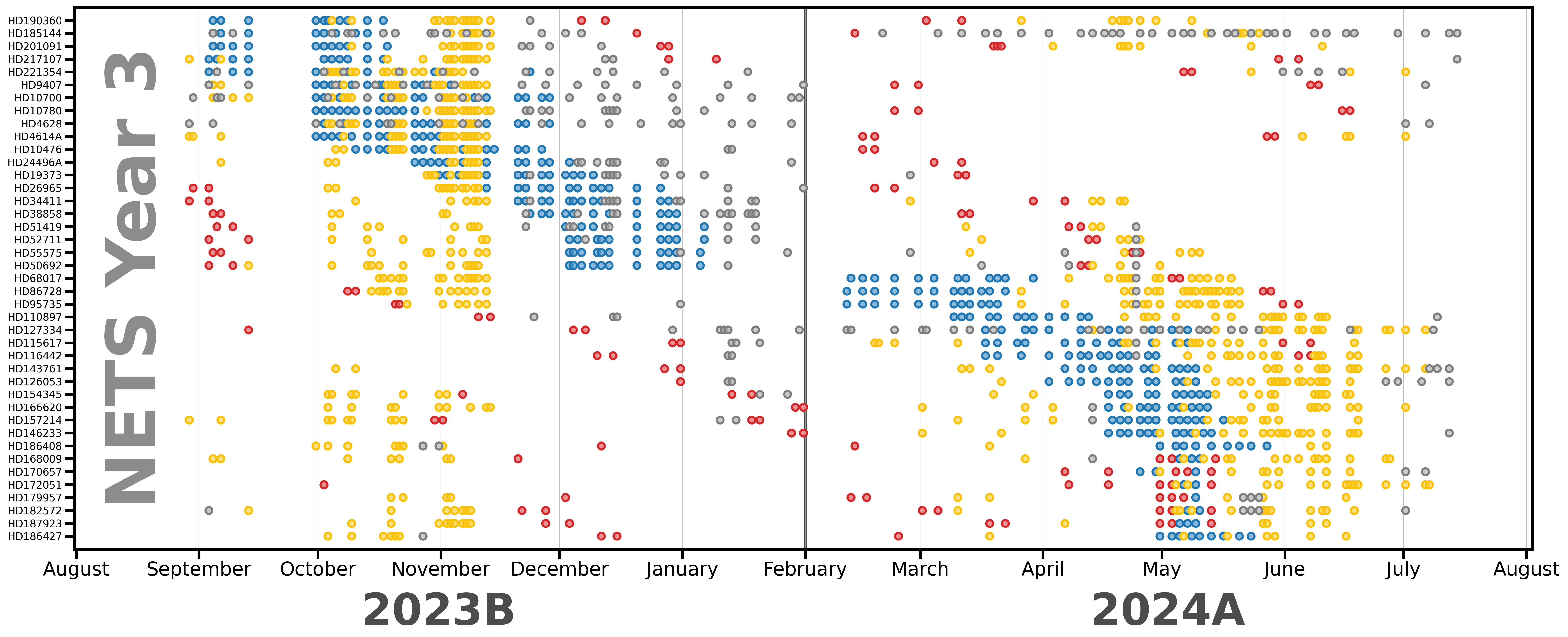}
    \caption{Top: \rev{Representative} seasonal observing schedule for NEID Earth Twin Survey target stars. We schedule sets of time-sensitive observations (red; P0) to bracket the observing season, high cadence observations (blue; P1) in the middle of each season to sample rotationally-modulated stellar signals, and long baseline observations (yellow; P2) to improve phase coverage. Lower panels: Enacted schedules for each of the first three years of the survey. In gray, we show standard star observations and supplemental NETS observations, including P3 and P4, that do not strictly follow our seasonal and nightly observing schemes.}
    \label{fig:nets_schedule}
\end{figure*}

For each star, we schedule P1 visits in the middle of each season at high cadence, set to execute as frequently as once every two nights. These observations, which typically span 1--2 months, allow us to densely sample the RV time series on timescales comparable to the rotation periods of our target stars. Rotationally-modulated stellar activity signals are expected to complicate exoplanet detection for even the quietest stars \citep{Crass2021}, but simultaneous measurement of RVs and activity indicators can be a powerful tool for disentangling stellar and planetary signals at the sub-m s$^{-1}$ level. However, this requires that our data sets are sufficiently well sampled so that we can measure and model activity variations on the relevant timescales  \citep{Rajpaul2015,Haywood2018,Hall2018,Lanza2019,Yu2024,Gupta2024}.
On their own, these densely-sampled observations would lead to poor phase sampling for longer period exoplanet signals, which can degrade the accuracy and precision of the extracted signal amplitude \citep{Burt2018}. To improve our phase sampling, we use our P2 allocation to schedule low cadence observations that span each star's observing season. Finally, to reduce annual aliasing effects (which can be a significant obstacle to the detection of planet with periods close to one year), we schedule sets of 2--3 P0 visits as each star is rising or setting to bracket the observing season. By making use of the highest priority time for these observations, we ensure that the stars are observed within a few weeks of each end of the observing season when they are observable for no more than a couple hours each night. This counteracts the tendency for targets to drop out of the schedule during these limited observability windows. A \rev{representative} version of the seasonal observing schedule for a typical star\rev{, with ideal uniform spacing between visits,} is shown in the top panel of \autoref{fig:nets_schedule}. The schedule is supplemented by additional P3 and P4 time provided by Penn State; these observations follow the same nightly observing strategy, but no constraints are imposed on the observing cadence at these low priorities. We allow each star to be observed no more than once per night across all priority levels.

\subsection{Current Status of NETS Observations}\label{sec:obs_status}

NETS observations formally commenced in September 2021, at the start of the 2021B observing semester. We show the achieved NETS observing schedule through August 2024 in \autoref{fig:nets_schedule}. In addition to the observations that follow the seasonal and nightly strategies outlined above, we also include supplemental data taken outside of this algorithmic scope and data from the NEID standard star program. The supplemental data adhere to the nightly NETS observing strategy (Section \ref{sec:single_night}) and these include observations at lower priorities (P3 and P4) as well as observations scheduled on an ad hoc basis to monitor interesting signals or to compensate for weather losses and other scheduling setbacks. The target list for the NEID standard star program includes the following NETS targets: HD\,4628, HD\,9407, HD\,10700, HD\,127334, HD\,185144, and HD\,221354. The standard star program uses the same S/N trigger as NETS for all but HD\,10700, for which a slightly lower threshold is used. For the analysis in this work, we combine data from both programs, as the RV uncertainty introduced by CTI variations is not expected to exceed 10 cm s$^{-1}$ for observations at high S/N \citep{Blackman2020}. For more detailed analyses, one could consider separating the NETS and standard star data for HD\,10700. 

Data collection proceeded smoothly through the 2021B semester but was interrupted twice during the 2022A semester. The first interruption resulted from a failure of the etalon calibration source in early May 2022. NETS observations were paused for three weeks until the light source was replaced and demonstrated to be stable. The second interruption began in June 2022 when NEID was shut down and allowed to warm up to ambient temperature to save the instrument from the observatory-wide power shutdown due to the Contreras fire. The instrument was restarted and cooled to its operating temperature in October 2022, NEID returned to on-sky operations on November 11, and NETS observations recommenced on November 30.
Observations during the 2022B semester (shortened to November 2022 through January 2023) did not adhere to our seasonal strategy, as our top priority was simply to maximize data collection to make up for lost time. Nearly all stars were made perpetually available to the queue on every priority level. We returned to the pre-restart algorithmic scheme at the start of the 2023A semester and observations again proceeded until the WIYN primary mirror support system failed in September 2023, taking the telescope offline for two weeks. Observations resumed at the end of the month and there were no further interruptions through the end of the 2024A semester. We note that standard star observations were taken  during these first two interruptions (the etalon outage and as soon as NEID was back on sky following the Contreras fire) to continue monitoring the state of the instrument. Though we include these in Figure \ref{fig:nets_schedule},  they are not included in any analysis in this work, as the stability of the instrument in these intervals would not necessarily be sufficient to deliver the high RV precision our survey requires.

The observing schedules shown in Figure \ref{fig:nets_schedule} also exclude data that were identified as problematic for one or more of the following reasons:
\begin{itemize}
    \item \textbf{Low S/N:} NETS observations are set to terminate at a maximum exposure time equal to twice the nominal value if they have not already reached the specified S/N threshold. In exceptionally poor observing conditions, this will happen while the S/N is still very low. We exclude data with S/N $\leq60$\% of the requested value.
    \item \textbf{Wrong target observed:} Our target sample includes several visual binaries. In some cases, the wrong star in each binary pair has been observed due to acquisition errors. We identify and exclude these observations using the known difference between the systemic velocities of each star, which is orders of magnitude greater than the observed night-to-night RV variability. The affected stars are HD 116442, HD 179957, HD 186408, and HD 186427.
\end{itemize}

The total number of nights of high quality observations for each NETS target is given in \autoref{tab:nets_stats}.
While some stars reached the intended rate of 36 visits per year, many others did not. This shortfall is a consequence of multiple factors including weather losses, which are most pronounced in February and July, as well as the aforementioned facility interruptions. The stars that were most affected are those whose observing seasons overlap substantially with these poor weather windows, the interruptions, and the annual observatory shutdown each August. Still, as we show in \autoref{fig:nets_schedule}, we did manage to schedule and execute the full set of densely sampled P1 observations each year for nearly every star, with the exception of those planned for the truncated 2022B semester.

\subsection{Data Products}\label{sec:data_products}

This work uses NETS data as processed by version 1.4 of the NEID Data Reduction Pipeline (DRP)\footnote{Pipeline documentation can be found at \url{https://neid.ipac.caltech.edu/docs/NEID-DRP/}}, which uses the cross-correlation function \citep[CCF;][]{Baranne1996} method with spectral type-dependent masks to calculate RVs. We retrieved the processed spectra from the NEID data archive.\footnote{\url{https://neid.ipac.caltech.edu}} Data collected during Science Run 1 (2020 October 26 -- 2022 June 16) and Science Run 2 (2022 October 18 -- 2024 August 19) are treated as distinct RV eras, as the thermal cycling of the spectrograph in response to the Contreras fire is expected to have introduced an offset in the instrumental RV zero point\footnote{\url{https://neid.ipac.caltech.edu/docs/NEID-DRP/rveras.html}}. We have also identified an apparent RV zero point offset in the middle of Science Run 1, the physical cause of which is yet unknown, so we consider data taken from 2020 October 26 -- 2021 August 31 as a separate ``Run 0.5''.
We refer to these as Run 0.5, Run 1, and Run 2 in this work, and the offsets are described in detail in Section \ref{sec:zeropoint}. 

\begin{figure*}
    \centering
    \includegraphics[width=\linewidth]{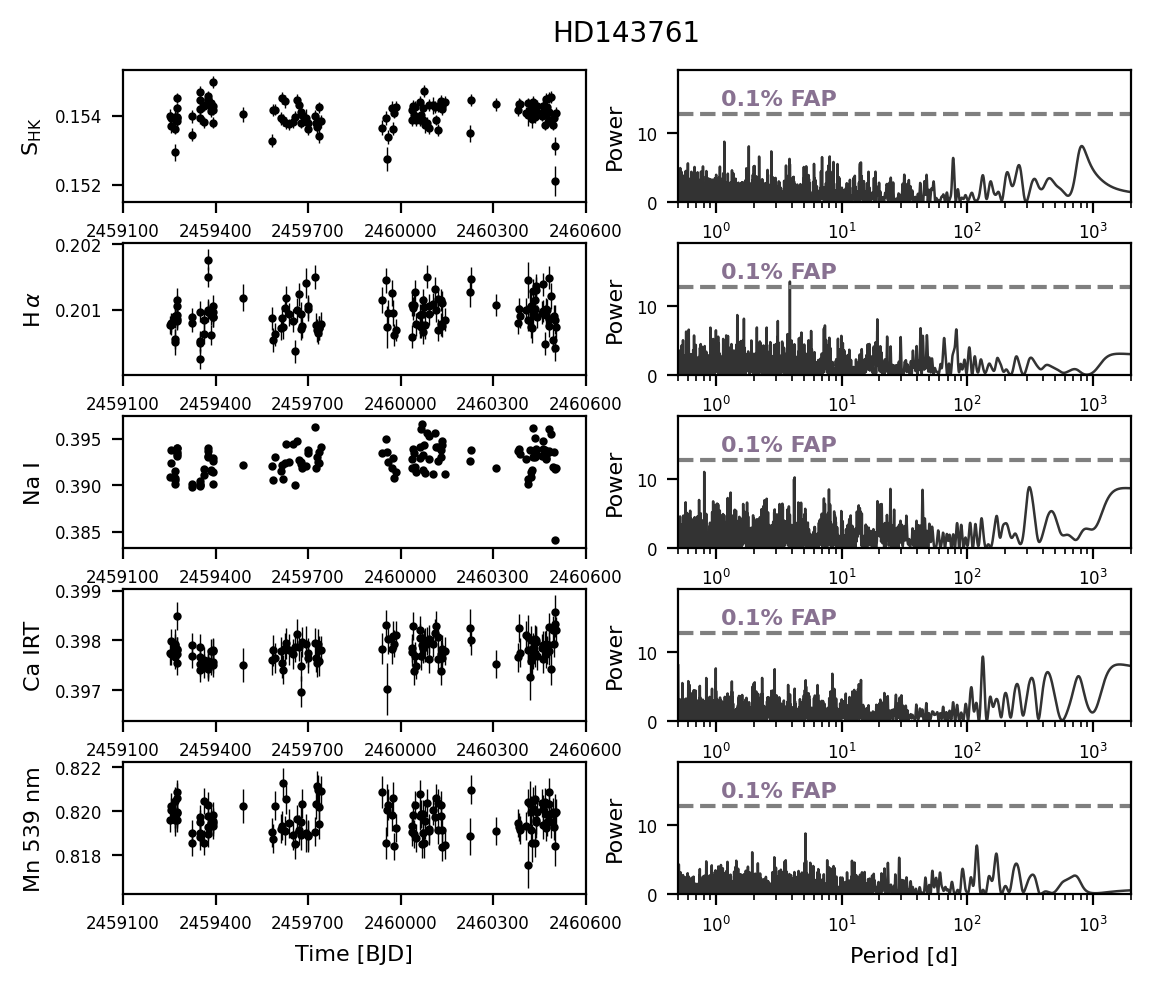}
    \caption{NEID activity indicator time series measurements and GLS periodograms for HD\,143761. We adopt a 0.1\% FAP threshold to evaluate the significance of any detected signals; this is shown as a horizontal dashed line in each panel. We do not show the FWHM or BIS, the interpretation of which would be clouded by the line profile changes described in Section \ref{sec:zeropoint}. For this star, no significant activity signals are detected. Equivalent figures for other targets will be provided as part of a figure set.}
    \label{fig:activity_timeseries}
\end{figure*}

In addition to precise RVs, the DRP provides useful diagnostics such as the CCF full width at half maximum (FWHM) and bisector inverse slope (BIS) as well as telluric-corrected measurements of various indicators of stellar activity levels. These are calculated as the ratio of the flux in the specified activity-sensitive line core relative to a nearby continuum region, as described in the DRP documentation. Among these indicators is the Ca II H\&K emission, though the values provided by the NEID DRP have not been calibrated to the standard Mount Wilson (MW) scale for the commonly used S$_{\rm HK}$ metric \citep[][ also called the S-index or S-value]{Duncan1991}. We follow the approach of \citet{Lovis2011} and \citet{GomesdaSilva2021} and compute the scaled S$_{\rm HK, MW}$ values from the native DRP Ca II H\&K activity indicator by performing a linear fit between the average Ca II H\&K values from the NEID DRP and the mean S-values from \citet{Duncan1991} for the 27 FGK stars that appear in both samples. We find a conversion S$_{\rm HK, MW} = 1.177$ Ca II H\&K$_{\rm DRP} + 0.004$, and we expect this calibration to be valid for stars with spectral types and activity levels comparable to the input sample (FGK stars with $0.12<$ Ca II H\&K$_{\rm DRP} <0.55$). We report both the DRP Ca II H\&K values as well as the scaled S$_{\rm HK, MW}$ for each star, with the exception of HD\,95735 \rev{(GJ\,411)}, the sole M-dwarf on the NETS target list. 
The time series and generalized Lomb-Scargle \citep[GLS;][]{Zechmeister2009,Zechmeister2018gls} periodograms for several activity-sensitive lines are shown in \autoref{fig:activity_timeseries} for a typical NETS target, and in the accompanying figure set for the remaining target stars. False alarm probabilities (FAP) are calculated following \citet{Zechmeister2009}, accounting for the number of independently sampled frequencies.

\input{FigSet2}

\section{Survey Diagnostics}\label{sec:diagnostics}

\subsection{Planetary companions}

Nine of the 41 stars in our sample are known to host planetary systems. These are HD\,86728, HD\,95735, HD\,115617, HD\,143761, HD\,154345, HD\,168009, HD\,186427, HD\,190360, and HD\,217107. We list each of these planets and the latest orbital parameter measurements from the literature in Table \ref{tab:lit_planets}, and these signals are discussed in more detail in Section \ref{sec:rvsearch}. Here, we calculate the residual NEID RV signal for each star by subtracting the published orbit model from the observed NEID RVs. These residual RV measurements are used for the remainder of the analysis in the following subsections.

\begin{deluxetable*}{lccccccc}
\tabletypesize{\scriptsize}
\tablecaption{Orbital Parameters for Known Planetary and Stellar Companions}
\label{tab:lit_planets}
\tablehead{
\colhead{Name} & \colhead{$P$ (d)} & \colhead{$e$} & \colhead{$\omega$ ($^{\circ}$)} &
\colhead{$K$ (m\,s$^{-1}$)} & \colhead{$T_{\rm conj, inf}$ (d)} & \colhead{$T_{\rm peri}$ (d)} &\colhead{Reference}
} 
\startdata
HD 143761 b & $39.8438 \pm 0.0027$ & $0.038 \pm 0.0025$ & $ -90.36^{+3.44}_{-3.32} $ & $ 67.28 \pm 0.19$ & $ 2455479.62 \pm 0.29 $ & $2455498.7 ^{+0.75}_{-39.0}$  & (1) \\ 
HD 143761 c & $102.19 ^{+0.27}_{-0.22}$ & $0.096 ^{+0.053}_{-0.054}$ & $9.7 ^{+25}_{-31}$ & $4.0 ^{+0.18}_{-0.17}$ & $2455629.3 ^{+8.6}_{-11.0}$ & $2455609 ^{+13}_{-14}$ & (1) \\ 
HD 143761 d & $282.2 ^{+2.2}_{-3.7}$ & $\equiv 0.0$ & $\equiv 0.0$ & $2.2 \pm 0.14$ & $2455583 ^{+54}_{-31}$ & $2455512 ^{+55}_{-32}$ & (1) \\ 
HD 143761 e & $12.949 \pm 0.014$ & $0.126 ^{+0.054}_{-0.078}$ & $-0.6 ^{+49.3}_{-57.3}$ & $1.14 \pm 0.015$ & $2455498.5 ^{+4.6}_{-4.7}$ & $2455496.2^{+5.7}_{-5.0}$ & (1) \\ 
HD 190360 b & $2892 \pm 4$ & $0.3184 ^{+0.0065}_{-0.0063}$ & $14.8 \pm 1.4$ & $20.25 \pm 10.43$ & \nodata & $2456428 \pm 10$ & (2) \\ 
HD 190360 c & $17.1169664 \pm 0.0005113$ & $0.165 ^{+0.023}_{-0.024}$ & $323.4 ^{+9.2}_{-9.1}$ & $4.542 \pm 4.071$ & \nodata & $2455210.27 \pm 0.42$ & (2) \\ 
HD 217107 b & $7.126870132 \pm 0.000004748$ & $0.1284 ^{+0.0015}_{-0.0014}$ & $22.78 ^{+0.71}_{-0.70}$ & $137.3 \pm 26.1$ & \nodata & $2455201.109 \pm 0.014$  & (2) \\ 
HD 217107 c & $5138.6 \pm 11.3$ & $0.3918 ^{+0.0064}_{-0.0067}$ & $205.0 \pm 1.1$ & $52.95 \pm 2.75$ & \nodata & $2455911 \pm 15$ & (2) \\ 
HD 154345 b & $3280 ^{+29}_{-28}$ & $0.058 \pm 0.019$ & $309 ^{+18}_{-21}$ & $16.70 \pm 1.96$ & \nodata & $2458230 ^{+147}_{-263}$ & (2) \\ 
HD 186427 b & $799.45 \pm 0.15$ & $0.6832 \pm 0.0031$ & $82.74 \pm 0.60$ & $53.16 ^{+0.32}_{-0.31}$ & $2456937.32 \pm 0.38$ & \nodata & (3) \\ 
HD 115617 b & $4.21487 \pm 0.00016$ & $0.0990 ^{+0.0910}_{-0.0700}$ & $20 ^{+48}_{-69}$ & $2.17 ^{+0.22}_{-0.21}$ & $2455200.70 ^{+0.14}_{-0.11}$ & \nodata & (3) \\ 
HD 115617 c & $38.0957 \pm 0.0083$ & $0.0680 ^{+0.0670}_{-0.0470}$ & $-20.0 ^{+86.0}_{-44.0}$ & $3.27 ^{+0.21}_{-0.22}$ & $2454922.48 ^{+0.69}_{-0.65}$ & \nodata & (3) \\ 
HD 115617 d & $123.2 \pm 0.2$ & $0.120 \pm 0.050$ & $253 ^{+29.01}_{-34.0}$ & $1.44 \pm 0.08$ & $2455602.99 ^{+11}_{-15}$ & \nodata & (4) \\ 
HD 168009 b & $15.1479 ^{+0.0035}_{-0.0037}$ & $0.120 ^{+0.110}_{-0.082}$ & $13.18 ^{+45.84}_{-63.03}$ & $2.53 ^{+0.31}_{-0.30}$ & $2455201.56 ^{+0.82}_{-0.77}$ & \nodata & (3) \\ 
HD 95735 b & $12.9394 ^{+0.0014}_{-0.0013}$ & $0.063 ^{+0.061}_{-0.043}$ & $-29 ^{+183}_{-120}$ & $1.381 ^{+0.093}_{-0.092}$ & $2456305.38  ^{+0.30}_{-0.28}$ & \nodata & (5) \\ 
HD 95735 c & $2946 ^{+210}_{-180}$ & $0.132 ^{+0.160}_{-0.091}$ & $63 ^{+57}_{-178}$ & $1.16 ^{+0.20}_{-0.19}$ & $2457205 ^{+98}_{-96}$ & \nodata & (5) \\ 
HD 86728 b & $31.1503 ^{+0.0062}_{-0.0066}$ & $0.064 ^{+0.065}_{-0.044}$ & $94 ^{+43}_{-123}$ & $1.91 ^{+0.11}_{-0.12}$ & $2459260.43 ^{+0.53}_{-0.48}$ & \nodata & (6) \\ 
HD 4614 B & $172467.51 \pm 401.77$ & $0.49416 \pm 0.0007$ & $88.34 \pm 0.25 $ & $ 1021.5 \pm 8.9 $ & \nodata & $2583650^{+421}_{-419}$ & (2) \\
HD 68017 B & $26114.84 ^{+1826.21}_{-1497.49}$ &	$0.348^{+0.019}_{-0.016}$ & $267.8\pm3.4$ & $214.6\pm7.6$ &  \nodata &	$2457722\pm99$ & (2) \\
HD 24496 B & $215127.83^{+20818.82}_{-30680.37}$	& $0.099^{+0.13}_{-0.054}$	& $318^{+13}_{-32}$	& $1330.8\pm 86.7$ &  \nodata 	 & $2557107^{+8299}_{-10043}$  & (2) \\
HD 201092 & $258262.97^{+1643.59}_{-1607.07}$ &	$0.4424	^{+0.0056}_{-0.0057}$ &	$330.61	^{+0.68}_{-0.67}$	& $1566.9\pm17.5 $ &  \nodata &	$2596496^{+1255}_{-1233}$ & (2) \\
\enddata
\tablenotetext{}{\textbf{Note.} Columns are orbital period ($P$), eccentricity ($e$), argument of periastron ($\omega$), RV semi-amplitude ($K$), time of inferior conjunction ($T_{\rm conj,inf}$), and time of periastron ($T_{\rm peri}$). References: (1) \citet{Brewer2023}, (2) \citet{Giovinazzi2025}, (3) \citet{Rosenthal2021}, (4) \citet{Cretignier2023}, (5) \citet{Hurt2022}, (6) \citet{Gupta2025}.}
\end{deluxetable*}

\subsection{Stellar companions}

Several of the stars in our sample are members of binary (or higher multiplicity) systems. To determine which of these stellar companions are expected to induce measurable RV signals, we use \texttt{binary\_mc} \citep{zenodo15352084} to simulate possible orbits based on the positions of each star as given by the Gaia DR3 catalog \citep{GaiaCollaboration2023}. We calculate the maximum RV slope for each of 10 million random orbit draws for each system, and we consider the slope to be measurable in systems for which at least 10\% of orbits have maximum slopes $\geq 1$ m~s$^{-1}$~yr$^{-1}$.
Four of these, HD\,4614\,A, HD\,24496\,A, HD\,68017, and HD\,201091 (61\,Cyg\,A), were analyzed in a study of NETS stars with high astrometric accelerations by \citet{Giovinazzi2025} which combined astrometric measurements, archival RVs, and Run 1 NETS data to refine the binary orbits. We adopt and subtract the published orbits from the observed RVs for each of these stars. The orbital parameters are given in Table \ref{tab:lit_planets}. An additional three stars, HD\,179957, HD\,186408, and HD\,116442 have measurable RV signals due to their stellar companions but did not meet the criteria for thorough characterization in \citet{Giovinazzi2025}. For these stars, we fit and subtract separate linear trends from NEID time series for each run.

Other stars in our sample with stellar-mass companions include HD\,86728, which has a low-mass M-dwarf binary \citep{Gizis2000,Hirsch2021} and HD\,26965, which is part of a triple star system in which the orbit of the B and C system is well characterized \citep{Mason2017} but that of the BC pair about the primary star is not. For these systems, we neither expect nor observe measurable RV signals as a consequence of some combination of long orbital periods, low companion masses, and highly inclined orbits.

\subsection{RV Zero Point Offset}\label{sec:zeropoint}

\begin{figure}
    \centering
    \includegraphics[width=\linewidth]{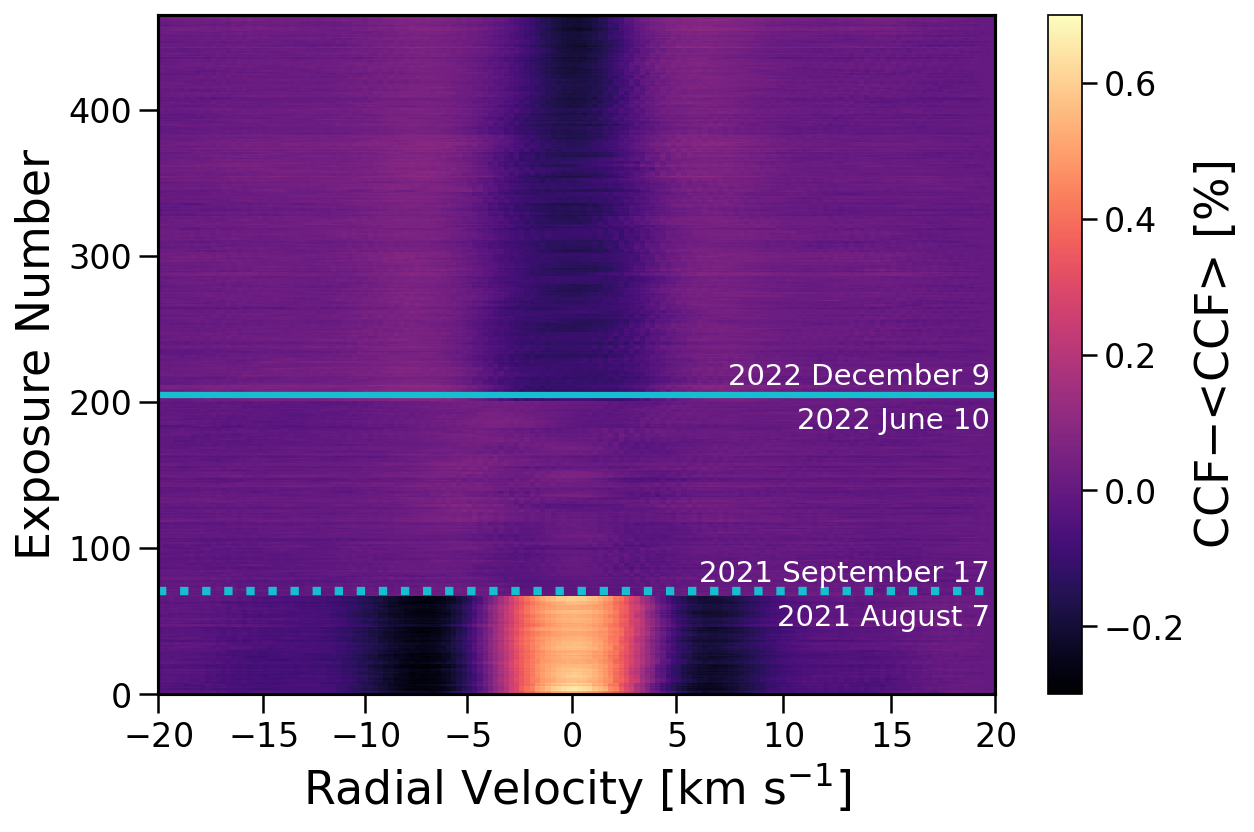}
    \caption{Residual CCFs for HD\,185144. The average CCF profile exhibits two sharp transitions. The latter of these corresponds to the thermal cycling of NEID following the Contreras fire, but the earlier transition in Summer 2021 remains unexplained. The more gradual decrease in the depth of the CCF core during Run 2 is the result of a matching decrease in the stellar activity level over that time period.}
    \label{fig:ccf_colormap}
\end{figure}

\begin{figure}
    \centering
    \includegraphics[width=\linewidth]{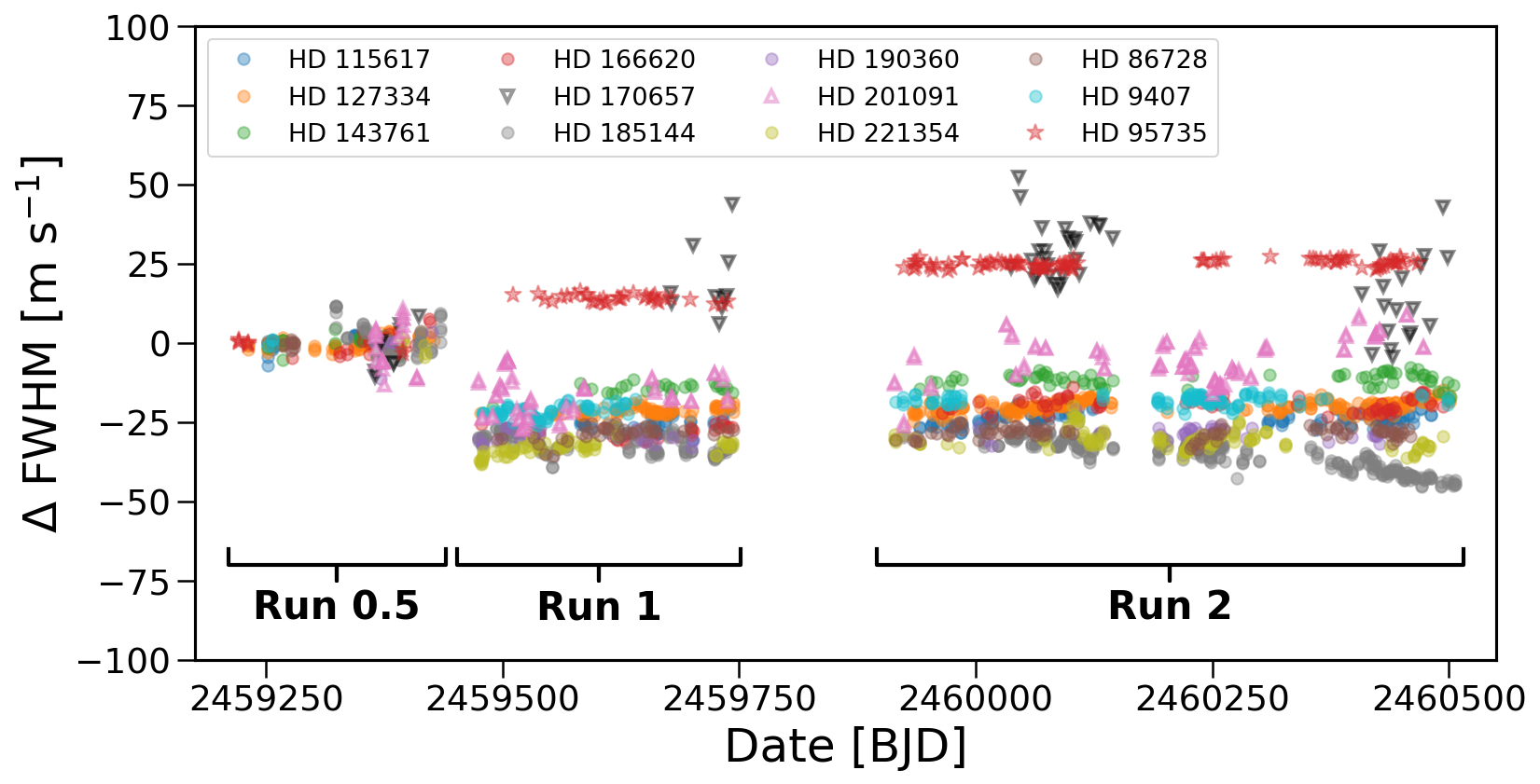}
    \caption{Time series of relative FWHM measurements for all NETS stars with multiple nights of data collected prior to August 2021. We observe a sharp decrease in FWHM in August 2021 for all Solar-type stars, and we interpret this as a break in the RV time series necessitating the definition of a new NEID RV era prior to this date. A sharp FWHM change is also observed for HD\,95735, an M-dwarf, but this change does not follow that of the other stars, suggesting a chromatic dependence. Though the $\Delta$FWHM time series for HD\,170657 and HD\,201091 stand out from the other Solar-type stars, these variations reflect changing stellar activity levels rather than an instrumental effect.}
    \label{fig:fwhm_change}
\end{figure}

The extraordinary stability of EPRV spectrographs such as NEID is achieved through careful environmental control with extremely strict tolerances \citep[e.g.,][]{Stefansson2016,Robertson2019,Crass2022}. As a consequence, changes to the reference wavelength scale and average line profile resulting from even minor changes to the spectrograph or its calibration sources may introduce a measurable RV zero point offset. Our analysis of the line profiles of NETS observations reveals that two such changes occurred during the first three years of the survey. In \autoref{fig:ccf_colormap}, we show the residual CCFs for all spectra of the star HD\,185144, which was observed as part of both the NEID standard star program and NETS. The residual CCFs are calculated as the difference between the \rev{normalized} CCF of each observation and the average \rev{normalized} CCF over all observations. There are two distinct changes to the average line profile; the first occurs between 2021 August 7 and 2021 September 17, and the second occurs between 2022 June 10 and 2022 December 9. The latter change corresponds to the break between Run 1 to Run 2 and is explained by the derivation of a new master wavelength solution for the Thorium-Argon (ThAr) calibration lamp and by the thermal cycling of the instrument in response to the Contreras fire. The earlier break corresponds to dates of the annual observatory shutdown in 2021, but there is no record of any hardware or software changes that would have produced such a drastic CCF shape change. We confirm that this is not isolated to HD\,185144 by computing the CCF FWHM time series for all stars that were observed during our precursor study in early 2021 (\autoref{fig:fwhm_change}). Nearly all stars see the same $\sim$ 25 m~s$^{-1}$ decrease in the average line width. We therefore find it necessary to define an additional RV era for data taken prior to 2021 August 31. We call this Run 0.5, electing to continue to use Run 1 to refer to data collected after this date and before the Contreras fire.

For CCF-based RV calculations, such as that of the NEID DRP, the RV zero point offsets resulting from these line profile changes will depend on additional factors such as projected rotational velocity ($v \sin i$) and effective temperature ($T_{\rm eff}$), which affect the intrinsic RV information content of the stellar spectrum \citep{Bouchy2001}, on the CCF mask, which dictates which lines are used to calculate the RV, and on the systemic velocity ($\gamma_\star$), which determines where these lines fall on the detector. Therefore, while the reference wavelength shifts are expected to be stable, the RV offset will vary from star to star.

\begin{figure}
    \centering
    \includegraphics[width=\linewidth]{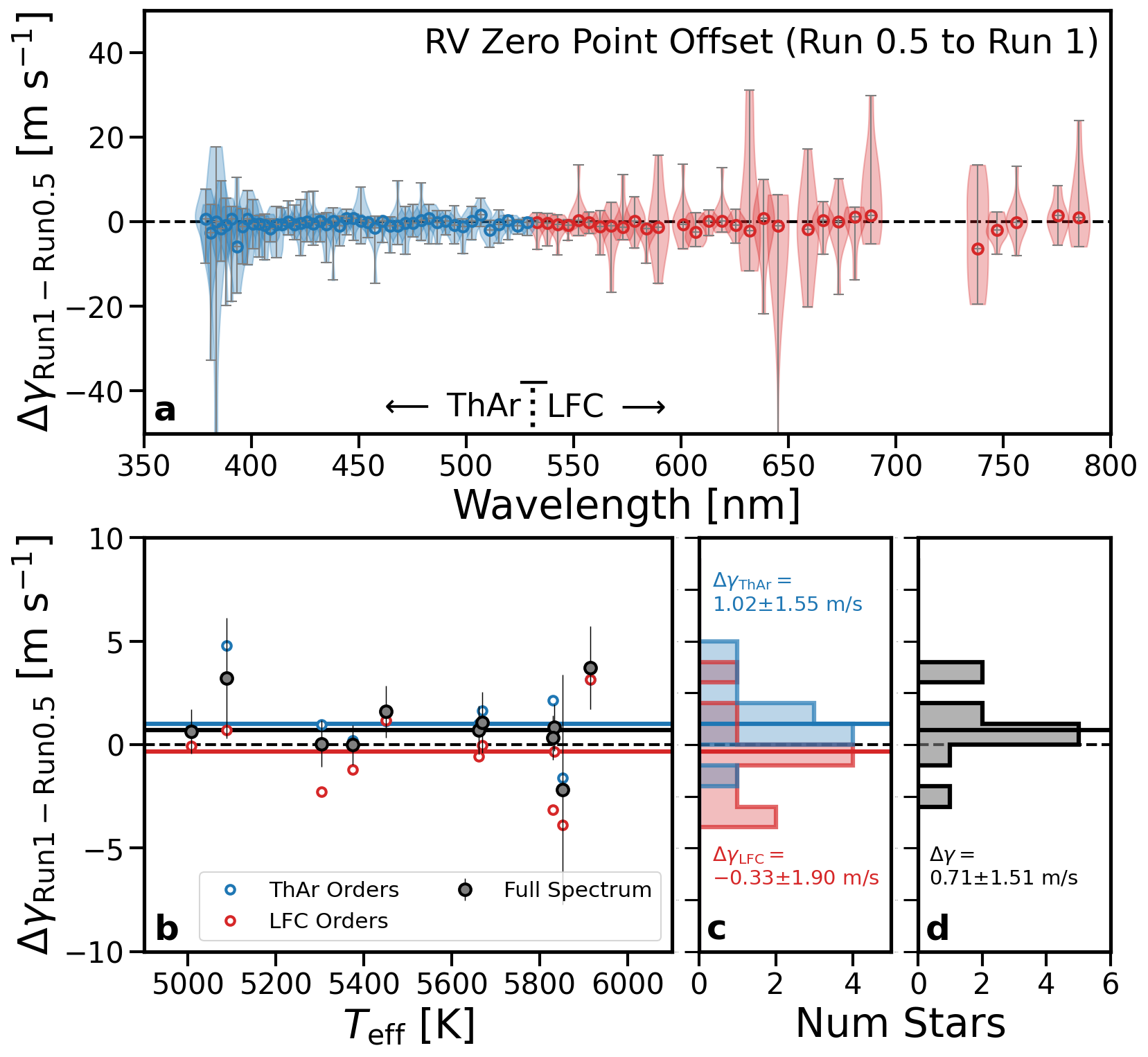}
    \caption{NEID RV zero point change between Run 0.5 and Run 1. We show the zero point change as a function of spectral echelle order (a) and as a function of stellar temperature (b). No significant chromatic trends emerge.}
    \label{fig:zeropoint_fwhm}
\end{figure}

\begin{figure}
    \centering
    \includegraphics[width=\linewidth]{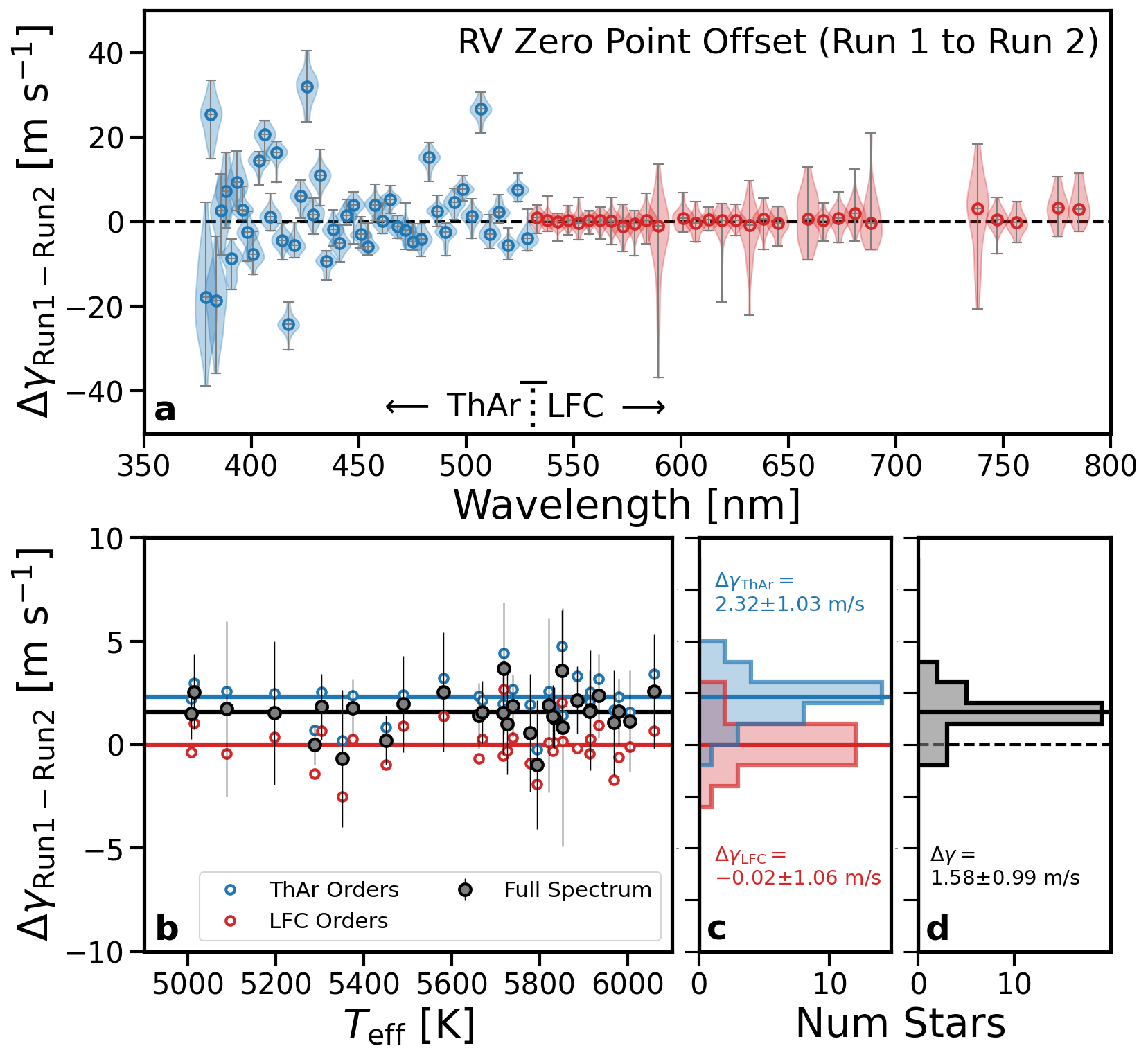}
    \caption{Same as \autoref{fig:zeropoint_fwhm} but for the RV break between Run 1 and Run 2. Significant order-to-order variations are present for orders with wavelength solutions anchored by the ThAr calibration lamp, while all LFC orders have a zero point change consistent with zero.}
    \label{fig:zeropoint}
\end{figure}

To quantify each of these two zero point changes across our target sample, we first subtract the RV signals of known stellar and substellar companions and then fit an offset between the Run 0.5 and Run 1 NETS RV time series (\autoref{fig:zeropoint_fwhm}) and the Run 1 and Run 2 NETS RV time series (\autoref{fig:zeropoint}) for each star. We exclude stars that have companions with orbital periods greater than the baseline of our survey to date, as moderate uncertainties in the orbital solutions could significantly bias the average residual RV across each run. The mean RV zero point offsets are determined to be 
$\Delta \gamma_{{\rm Run 1} - {\rm Run 0.5}}=0.71\pm1.51$ m~s$^{-1}$
and
$\Delta \gamma_{{\rm Run 1} - {\rm Run 2}}=1.58\pm0.99$ m~s$^{-1}$. The narrow distribution of values across the stellar sample is encouraging, as it allows strong priors to be imposed when fitting for exoplanet-induced RV signals using data sets that span multiple runs.

The above results were derived using late F-, G-, and early K-dwarfs, which make up the majority of the NETS sample. However, a chromatic (wavelength-dependent) zero point offset is hinted at by the FWHM time series for \rev{the lone M-dwarf in our sample,} HD\,95735 (\autoref{fig:fwhm_change}); this star was not used to calibrate the zero point offsets due to the presence of its long-period companion HD\,95735\,c \citep[$P=2496^{+210}_{-180}$ d;][]{Hurt2022}. The FWHM for most stars decreases from Run 0.5 to Run 1 then holds relatively steady, though with a small increase, through Run 2. HD\,95735, HD\,170657, and HD\,201091 deviate from this pattern. This can be explained by stellar activity in the case of HD\,170657 and HD\,201091, both of which exhibit increasing and highly variable chromospheric emission over this time frame. But activity levels remained stable for HD\,95735 while the FWHM increases from Run 0.5 to Run 1 and increases again from Run 1 to Run 2. As \rev{an M-dwarf}, HD\,95735 has a CCF dominated by lines at longer wavelengths than those of the FGK-dwarfs that make up the rest of the sample. \rev{Using the same subset of stars as before,} we check for any chromatic dependence by calculating the zero point offset as a function of wavelength, fitting the change in the mean stellar RV for each order over the wavelength range 380 nm -- 790 nm (echelle orders 162--78). For the Run 0.5 to Run 1 zero point offset (\autoref{fig:zeropoint_fwhm}), we show that our data are insufficient to measure any significant chromatic variation, as the offset for each order is consistent with zero. For the Run 1 to Run 2 offset, however, we do observe an interesting chromatic signal (\autoref{fig:zeropoint}). In particular, we find large offset variations from order to order in the wavelength region shorter than 530 nm, anchored by the ThAr calibration lamp. We attribute this to the new ThAr master wavelength solution. Further, the offset across all orders anchored by ThAr is 
$\Delta \gamma_{{\rm ThAr, Run 1} - {\rm Run 2}}=2.32\pm1.03$ m~s$^{-1}$
while the offset across the redder orders, anchored by the laser frequency comb (LFC), is
$\Delta \gamma_{{\rm LFC, Run 1} - {\rm Run 2}}=-0.02\pm1.06$ m~s$^{-1}$.
One should therefore exercise caution when using the mean offsets calculated for the full spectrum to inform analyses of M-dwarfs, which derive very little RV information content from the ThAr orders. Our results suggest redder stars should be insensitive to the Run 1 to Run 2 RV offset, though this needs to be confirmed via a similar analysis using a sample of M-dwarfs.

The RV zero point offsets described here are not expected to affect measurements of activity indicators, which do not contain enough spectral information to resolve line shape changes,  hence the treatment of the measurements presented in \autoref{fig:activity_timeseries} as unified time series. However, metrics that are computed from ensembles of lines spanning the full spectrum, such as the FWHM and BIS are likely to be affected by the same changes that produced the observed RV offsets.

\subsection{RMS}\label{sec:rms}

\begin{figure}
    \centering
    \includegraphics[width=\linewidth]{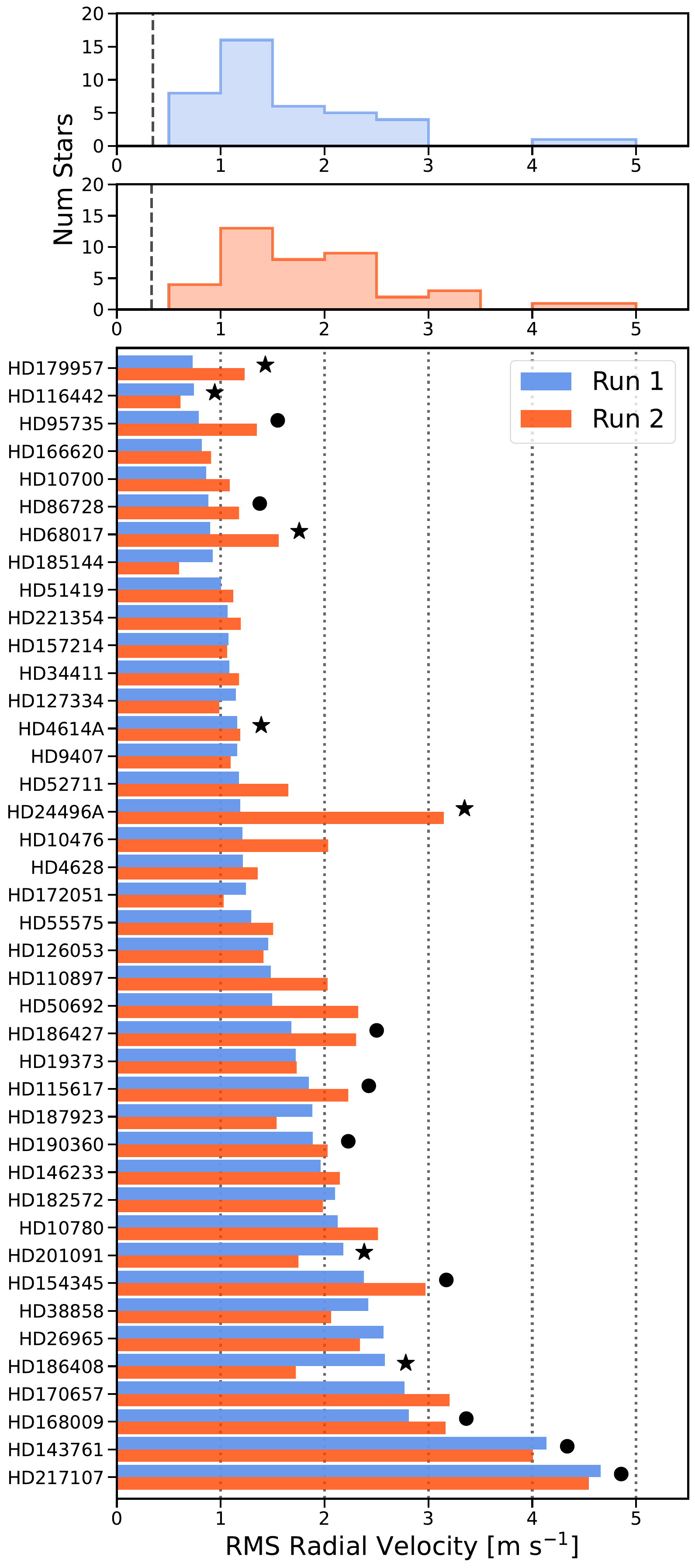}
    \caption{Run 1 (blue) and Run 2 (red) RMS RV for all NETS target stars after removing RV signals from known companions. We use $\star$ to indicate systems for which we subtracted signals from known stellar companions and $\bullet$ to indicate those for which we subtracted signals from known planetary companions. The majority of the stars have RMS RV between 1--2 m~s$^{-1}$. We also indicate the median single measurement precision as a black dashed line alongside the histograms for Run 1 and Run 2.}
    \label{fig:rms}
\end{figure}

As in Section \ref{sec:zeropoint}, we again subtract RV signals induced by known stellar and substellar companions, and we then calculate the nightly binned root mean square (RMS) RV for each star. These values are shown in \autoref{tab:nets_stats} and \autoref{fig:rms} for both Run 1 and Run 2; we omit Run 0.5 from this section due to the short baseline and limited stellar sample. Several stars exhibit sub-m~s$^{-1}$ scatter, and the majority of the NETS sample is variable at the 1--2 m~s$^{-1}$ level. 
The RMS consistently exceeds 2.5 m~s$^{-1}$ for very few of our targets; many of the higher RMS values correspond to stars with known planetary companions, suggesting either that the literature orbits need to be refined (i.e., large residuals in the NEID RVs result from a stale ephemeris), or that there are additional RV signals that we have failed to consider. The stars with known planets are discussed further in Section \ref{sec:rvsearch}. Here, we comment on some of the stars with large RMS but no known planets.

We measure RMS values of $>2$ m~s$^{-1}$ in both Run 1 and Run 2 for HD\,26965 and HD\,38858. As we note in Section \ref{sec:known_activity}, both stars exhibit moderate RV variations that were originally attributed to planetary companions but have since been determined to be driven by stellar activity. These variations are consistent with the elevated RMS we observe. The RMS value for HD\,24496\,A rises steeply from just $1.19$ m~s$^{-1}$ to $3.16$ m~s$^{-1}$ during Run 2. There is no coherent rotation signal in the RVs, but this increase is consistent with rising stellar activity levels as traced by the NEID S$_{\rm HK}$ and H$\alpha$ indicators, which in turn align with the 5.86-year activity cycle identified by \citet{Isaacson2024} using HIRES data.  HD\,170657 also appears highly variable, with the RMS just below 3 m~s$^{-1}$ in Run 1 and rising to just above 3 m~s$^{-1}$ in Run 2. No periodic signals are detected in the NEID RVs or activity indicators, but the relatively high S$_{\rm HK, MW}$ value of $0.328$ is consistent with elevated RV variability due to chromospheric activity \citep{Luhn2020}.

\section{\texttt{RVSearch} Analysis of NEID RVs}\label{sec:rvsearch}

We use \texttt{RVSearch} \citep{Rosenthal2021} to perform Bayesian Information Criterion (BIC) model comparisons via an iterative $\Delta$BIC periodogram search of the NEID RVs. The intent of this search is to probe the sensitivity of the NETS data set to known and new periodic signals for each target. In particular, we aim to test whether known planets can be independently recovered with NEID without reference to archival measurements. We therefore limit the analysis to NEID RVs, even though the inclusion of other RV data should in principle lead to a higher recovery rate. Below, we summarize key features of \texttt{RVSearch} that are relevant to our implementation of the algorithm.

The periodogram search is performed over a grid with the same $\frac{1}{2\pi\tau}$ frequency spacing used in \citet{Rosenthal2021}, where $\tau$ is the total observing baseline for each target.
A $\Delta$BIC periodogram is constructed by stepping through the period grid, fitting a fixed-period sinusoid to the data, and performing a $\Delta$BIC comparison between the $n+1$ and $n$ planet models.
As in \citet{Rosenthal2021}, we adopt a $\Delta$BIC detection threshold corresponding to an empirical false alarm level of 0.1\%.
At each step in the iterative planet search, if any periodogram peaks exceed this threshold, the most significant of these is used to perform a maximum \textit{a posteriori} Keplerian fit with all model parameters free. After the solution is determined, an additional planet is added to the model and another iteration of the periodogram search is carried out to compare the $n+1$ and $n$ planet models until no more peaks exceed the false-alarm threshold. For each step, the parameters for the maximum likelihood fit of each planet model are left free so that they may converge to an optimal solution as additional planets are identified. The NEID data from Runs 0.5, 1, and 2 are treated as separate time series for the purposes of this analysis, and we allow the model to freely fit the RV zero points for each run.

\texttt{RVSearch} can be initialized as a ``blind'' search, using a 1-planet Keplerian model with undefined parameters, or as an ``informed'' search, in which additional trend, curvature, and/or Keplerian terms are included to account for the RV signals from known companions. In the case of a blind search, a $\Delta$BIC goodness-of-fit comparison can be performed prior to the search to determine if linear or quadratic trend terms should be added to the model. We use varying combinations of this functionality for separate search cases, which we define as follows: stars with companions that have periods comparable to or longer than the observing baseline ($P>\tau/2$), and stars with no known companions or whose only companions have periods shorter than half the observing baseline ($P\leq\tau/2$). A flat prior is placed on the orbital eccentricity in all cases.

The maximum likelihood orbital parameters for the detected signals are given in Tables \ref{tab:rvsearch_results_trend} and \ref{tab:rvsearch_results_no_trend} as a means to compare them to the orbits of known planets. We stress that these should not be interpreted as updated measurements of the planets' orbits, as our approach omits archival RVs and a careful treatment of stellar variability, and we refrain from tailoring the analysis to any individual targets in favor of a uniform methodology for all systems.

\subsection{Stars with no known companions or only short-period companions}
Of the 41 stars in the NETS target sample, 25 have no known stellar or substellar companions and four host only short-period ($P\leq\tau/2$) companions. For these targets, our sensitivity to known signals should be limited by the signal amplitudes, the sampling cadence, and the noise properties of the observations, but not by the total observing baseline.

In these cases we perform two searches, both of which are blind to the orbital parameters of any known short-period companions. For the first search, we allow \texttt{RVSearch} to conduct the goodness-of-fit test for linear/quadratic trend terms and set the lower bound of the period grid to 2 days and the upper bound to $\tau$, the observational baseline of that star. For the second search, we do not test for trend terms but instead decrease the upper bound of the search grid to $\tau/2$. These separate tests are useful because instrumental offsets between the various NEID RV eras can be misinterpreted as a trend. This can be alleviated by decreasing model flexibility. The maximum  period is set to half the baseline for the second search because spurious signals with periods greater than $\tau/2$ are more frequently identified by \texttt{RVSearch} when no trend model is included. 

\subsection{Stars with known long-period companions}
The remaining 12 stars have known stellar or substellar companions with periods that exceed $\tau/2$ (seven with stellar companions, four with substellar/planetary companions, one with both a stellar and substellar companion). We consider these targets separately as our sensitivity to their long-period companions is likely impacted by the limited duration of the survey thus far.

When an orbital solution for the long-period companion is known, we perform two tests. In the first, we do not provide the orbital parameters of the known long-period companion but do allow for the goodness-of-fit test for trend/curvature. In the second, we provide the median literature values for the orbital parameters (given in Table \ref{tab:lit_planets}) and hold them fixed for the duration of the iterative planet search, such that \texttt{RVSearch} will not alter them even if another signal is identified. We do not provide the orbital parameters of companions with periods shorter than $\tau/2$. If a companion orbit is not known, we only run the goodness-of-fit test for linear/quadratic trend terms

In the special case of HD 186427 (16 Cygni B), which has both a stellar and planetary companion, the orbit of the planet is known but that of the stellar companion is not. We perform the two tests as above but when we inform the fit of the companion parameters, we only provide them for the planet.

\begin{figure}
    \centering
    \includegraphics[width=\linewidth]{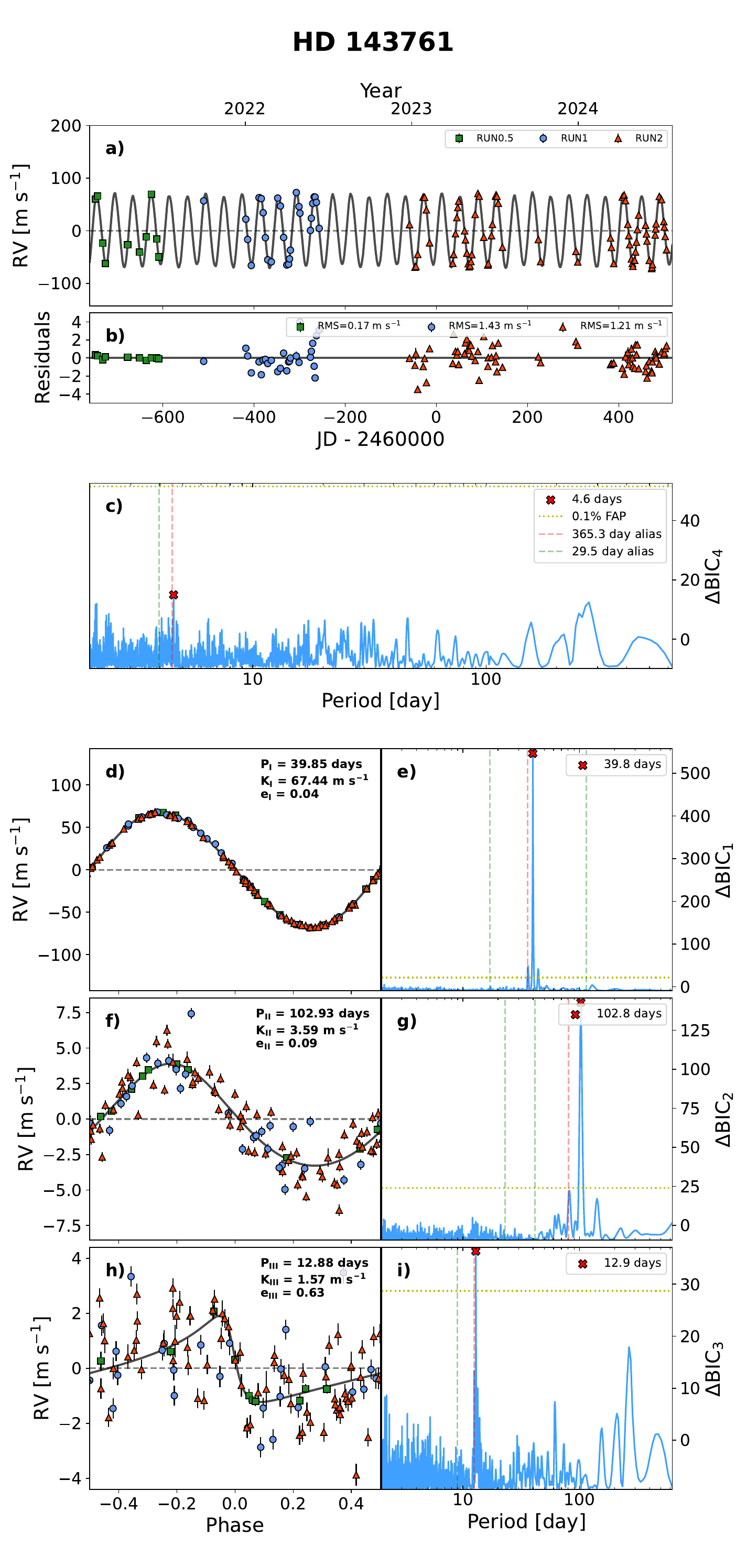}
    \caption{\texttt{RVSearch} results for HD\,143761 for the case in which no goodness-of-fit trend test was allowed and the period search was restricted to $P<\tau/2$. (a) Full orbit model and NEID RV time series, (b) residuals to the orbit model, (c) $\Delta$BIC periodogram of the RV residuals, (d) phase-folded orbit and (e) periodogram of the first detected signal, which corresponds to planet b, (f) phase-folded orbit and (g) periodogram of the next strongest signal, planet c, after removal of the first signal, and (h) phase-folded orbit and (i) periodogram of the final detected signal, planet e, after removal of the first two signals. Three signals were detected, corresponding to the known planets HD\,143761\,b, c, and e. Planet d is not detected, but it does appear among the strongest residual peaks. The complete figure set includes results for the \texttt{RVSearch} run with no trend for all stars with no known signals from long-period companions.}
    \label{fig:HD143761_rvsearch}
\end{figure}

In each the above cases, the bounds of the period search grid are set to 2 d $\leq P\leq\tau$. The \texttt{RVSearch} results for each system are shown in Figures \ref{fig:HD143761_rvsearch} (no known long-period companions, no trend), \ref{fig:HD190360_rvsearch_blind} (known long-period companions, blind), \ref{fig:HD190360_rvsearch_informed} (known long-period companions, informed), and \ref{fig:HD146233_rvsearch} (no known long-period companions, trend allowed) and their accompanying figure sets.

\section{Detected Signals}\label{sec:known_signals}

\subsection{Known Planet Systems - Short Periods}
There are four planet hosts in our sample for which all known companions have $P\leq\tau/2$. These are HD\,86728 (1 planet), HD\,115617 (\rev{3} planets), HD\,143761 (4 planets), and HD\,168009 (1 planet). For HD\,168009, we find that the goodness-of-fit test prefers a linear trend model, while the tests for the remaining stars prefer models without any trend/curvature terms. We find that, depending on the search parameters, either six or seven of these \rev{nine} known planet signals can be independently detected in the NETS data set.
HD\,168009\,b is detected when a linear trend is included in the fit, but the signal falls just short of the 0.1\% false alarm threshold in the $\Delta$BIC periodogram when no trend is included.
HD\,143761\,d \rev{and HD\,115617\,d} are not detected, \rev{though there is some disagreement in the literature as to whether this latter planet is real \citep{Rosenthal2021,Cretignier2023}}. All signals identified by the search that allows for a trend are listed in Table \ref{tab:rvsearch_results_trend}, and signals identified by the search without a trend are listed in Table \ref{tab:rvsearch_results_no_trend}.

\subsubsection{HD 86728}
HD\,86728 (GJ\,376\,A) is a slightly evolved G-type star with a wide separation M-dwarf companion, GJ\,376\,B \citep{Gizis2000}.
A 31-day RV signal was first reported by \citet{Hirsch2021} in an analysis of  measurements from the HIRES, APF, and the Lick-Hamilton spectrographs. The NEID data presented herein were later used by \citet{Gupta2025} alongside archival APF data to confirm this candidate signal. \citet{Gupta2025} find that HD\,86728\,b is an exoplanet with orbital period $P=31.1503^{+0.0062}_{-0.0066}$ d, projected mass $m_p\sin (i) = 9.16^{+0.55}_{-0.56}\ \rm{M}_\oplus$, and an orbital eccentricity consistent with zero.

The signal of HD 86728 b is identified in both of our searches, and the recovered parameters are consistent with the values given by \citet{Gupta2025}. No other prominent signals are identified.

\subsubsection{HD 115617}
The HD\,115617 (61 Vir)  system consists of a G-dwarf orbited by \rev{3} planets first discovered by \citet{Vogt2010}. HD\,115617\,b and c are super-Earth and Neptune-mass planets on nearly circular 4- and 38-day orbits, respectively \citep{Rosenthal2021}.
\citet{Vogt2010} reported a third planet, HD\,115617\,d, at a period of 124 days, \rev{which was also independently confirmed by \citet{Laliotis2023} and \citet{Cretignier2023}. However,} \citet{Rosenthal2021} noted that this signal is accompanied by harmonics at $\sim90$ days and residual power at 1-year, and following additional analysis of the correlation between the HIRES RVs and PSF parameters, they determined that the signal was a yearly systematic. 

Our searches return four signals in both cases. The first two signals are consistent with the orbital parameters of HD\,115617\,b and c. The other two signals are highly eccentric with periods of $\sim$92 and $\sim$390 days. The periodograms also show significant peaks at $\sim$120 and $\sim$180 days. These signals are all likely aliases \rev{of the 124-day orbit of HD\,115617\,d, but it is not clear whether this is a false positive, as suggested by \citet{Rosenthal2021}, or a real planet signal.}

\subsubsection{HD 143761}
HD\,143761 ($\rho$ CrB), is a near solar mass G-type subgiant with a low metallicity ([Fe/H] = -0.2) and isochrone derived age of $10.2 \pm 0.5$ Gyrs \citep{Brewer2023}. The system contains four confirmed planets. 
HD\,143761\,b is a 40-day Jupiter-mass planet and was one of the first giant planets discovered with RVs \citep{Noyes1997}. HD\,143761\,c was subsequently discovered by \citet{Fulton2016} with an orbital period of 102 days, and \citet{Brewer2023} added planets d and e which have 282-day and 13-day orbits, respectively. 

In both of our searches, three signals were identified and the best-fit orbital parameters are consistent with those of HD\,143761\,b, c, and e, though we measure a significantly higher eccentricity for HD\,143761\,e than that reported by \citet{Brewer2023}. The searches do not detect the 282-day period of HD\,143761\,d; the signal is present in the $\Delta$BIC periodograms of the residuals for both searches but it is well below the 0.1\% false alarm threshold.

\input{FigSet8}

\subsubsection{HD 168009}
HD\,168009 is a G-dwarf at a distance of 23.3 parsecs hosting a super-Earth discovered via RVs and first reported in \citet{Rosenthal2021}. They reported minimum mass of $m_p\sin (i)=9.53^{+1.21}_{-1.18}$ M$_{\oplus}$, an orbital period of 15.1479 $^{+0.0035}_{-0.0037}$ days and eccentricity of 0.121$^{+0.110}_{-0.082}$. They also report a 1-year periodic signal that is correlated with instrument systematics and is deemed a false positive.

Our search with a linear trend of $-3.0$ m~s$^{-1}$~yr$^{-1}$ recovers a 15.1-day signal consistent with the orbit of HD\,168009\,b as well as a moderately eccentric 61-day signal with no published counterpart.
Our search without a trend returns no identified signals in the RV time series. The highest peak in the periodograms correspond to a signal at a period of 15.1 days, consistent with that of HD\,168009\,b. However the peak does not exceed the false-alarm threshold and is thus not formally detected; the threshold is $\Delta$BIC $=18.7$ while the peak only reaches $\Delta$BIC $=18.6$ (FAP$=0.11\%$) .

\subsection{Known Planet Systems - Long Periods}

Five of the stars in our sample host at least one planet with $P>\tau/2$. These are HD\,95735 (2 planets), HD\,154345 (1 planet), HD\,186427 (1 planet), HD\,190360 (2 planets), and HD\,217107 (2 planets). 
We do not anticipate detecting the five planets with periods that exceed the limit of our search grids but we investigate their impact on the results of performing both blind and informed periodogram searches. 
The blind searches do not recover any of these planets, as expected. In some cases, the signals exhibit power at a shorter period alias and lead to spurious detections.
Still, in both the blind and informed searches, we detect all three of the shorter period planets along with the planet candidate HD\,190360\,d, which was recently identified by \citet{Giovinazzi2025}. Table \ref{tab:rvsearch_results_trend} lists the signals identified by the blind searches, which allows for a trend, and Table \ref{tab:rvsearch_results_no_trend} lists the results of the informed searches, in which known long-period companion orbits are removed using pre-defined models. Each system and the results of our search are described in detail below.

\subsubsection{HD 95735}
HD\,95735 (GJ\,411) is the lone M-dwarf in the NETS sample. As one of the brightest and nearest M-dwarfs ($\sim$2.5 pc), HD\,95735 has a long history as a target for precision RV surveys. An inner planet, HD\,95735\,b was first discovered by \citet{Diaz2019} at an orbital period of 13 days, and a longer-period candidate, HD\,95735\,c was later identified by \citet{Rosenthal2021}, though with substantial period uncertainty. A more detailed analysis of the complete set of RVs for this target was carried out by \citet{Hurt2022}, who refined the orbital parameters for both planets and confirmed HD\,95735\,c with an orbital period of 8 yr. Additionally, \citet{Hurt2022} provide constraints on a slightly eccentric candidate signal at $\sim$215 days and argue that the planetary case cannot be ruled out despite the classification of this signal as an instrumental systematic by \citet{Rosenthal2021}.

In our blind search, a model without any trend/curvature is preferred and three signals are detected. The first has a period of 12.9 days and is consistent with the orbit of HD\,95735\,b.
Eccentric signals at 220 and 654 days are also detected. The first of these is close to the 215-day candidate signal reported by \citet{Hurt2022}, but the second has no known comparable signals.
Our informed search returns only one signal, which is again the 12.9-day signal corresponding to HD 95735 b. The longer period eccentric signals from the blind search are not identified, suggesting that these were spurious detections caused by the long term trend induced by HD\,95735\,c. 

We categorize HD\,95735\,b as a successful detection.
Evidence for the 215-day candidate signal is inconclusive, as it was absent from the informed search despite a possible detection in the blind search.

\subsubsection{HD 154345}
HD\,154345 is a nearby G-dwarf with a single long-period giant planet companion. Using RV measurements that did not span an entire orbit, \citet{Wright2007} placed initial constraints on a substellar companion, HD 154345 b. The planet was then confirmed by \citet{Wright2008},  with a 9-yr orbital period and a minimum mass of close to 1 M$_{\rm Jup}$. The nature of the RV signal was later called into question due to a strong correlation between the RVs and the stellar activity level \citep{Wright2016}, but several follow up studies have maintained the RV variations to be of planetary origin \citep{Rosenthal2021,Xiao2023}. 
Most recently, \citet{Giovinazzi2025} jointly fit archival RVs and Run 1 NETS RVs with astrometric measurements from Hipparcos and Gaia to constrain the 3-D orbital architecture of the system. The astrometry rules out a face-on orbit for HD\,154345\,b.  
\citet{Giovinazzi2025} report a dynamical mass constraint of ${1.186}_{-0.059}^{+0.095}$ $M_{\rm{Jup}}$, an orbital period of ${8.981}_{-0.076}^{+0.079}$ years, and an eccentricity of ${0.058}\pm{0.019}$.

Our blind test opts for a linear trend term of $-4.8$ m~s$^{-1}$~yr$^{-1}$, which effectively accounts for the signal from HD\,154345\,b over the NEID baseline. While no signals exceed the false-alarm threshold, there are sharp peaks at 11 and 26 days in the $\Delta$BIC periodogram. The results for the informed search are similar; after subtracting HD\,154345\,b, periodogram peaks emerge at 11 and 26 days but no signals exceed the false-alarm threshold.

\subsubsection{HD 186427}
HD\,186427 (16\,Cyg\,B) is a G-dwarf with both an equal-mass visual binary companion, HD\,186408 (16\,Cyg\,A; also a NETS target), and a Jovian planet companion on a long-period (799-day), highly eccentric ($e=0.68$) orbit \citep{Cochran1997}.

The blind search prefers a second order trend model with a linear term of $-66.0$ m~s$^{-1}$~yr$^{-1}$ and a quadratic term of $0.9$ m~s$^{-1}$~yr$^{-2}$. We do not detect HD\,186427\,b, but one signal with a period of 370 days is identified.
However, this detection should be treated with scrutiny, as the fitted zero point offset between the Run 1 and Run 2 NEID RVs is $-120$ m~s$^{-1}$, considerably larger than expected based on the values derived in Section \ref{sec:zeropoint}.
In addition, the informed search returns no significant signals and no long-period power, so it is likely that the 370-day signal was produced by a poor fit to the orbit of HD\,186427\,b.

\subsubsection{HD 190360}
HD\,190360 is a mid-G type star with near-solar mass, temperature, and surface gravity. The system contains two confirmed planets with orbital periods of 2892 days \citep[HD\,190360\,b; first discovered by][]{Naef2003} and 17 days \citep[HD\,190360\,c; first discovered by][]{Vogt2005}. HD\,190360\,b is a Jupiter analog, with a true dynamical mass measurement of $m={1.68}_{-0.16}^{+0.26}~M_{\rm{Jup}}$ based on a combination of astrometric and RV measurements, including Run 1 NEID data \citep{Giovinazzi2025}. The astrometric signal of HD\,190360\,c is below the precision and baseline of the astrometric acceleration measurements, leading to a poor constraint on the mutual inclination of the system. \citet{Giovinazzi2025} also report a candidate signal in the NEID RV residuals with a period of 90 days; this signal had previously been identified but was attributed to a systematic alias by \citet{Rosenthal2021}.

The blind search adopts a second order trend model with a linear term of $-18.3$ m~s$^{-1}$~yr$^{-1}$ and a quadratic term of $6.6$ m~s$^{-1}$~yr$^{-2}$. Two signals are identified, one of which is consistent with the 17-day orbit of HD\,190360\,c, and the other is a nearly circular 89-day signal.

\begin{figure}
    \centering
    \includegraphics[width=\linewidth]{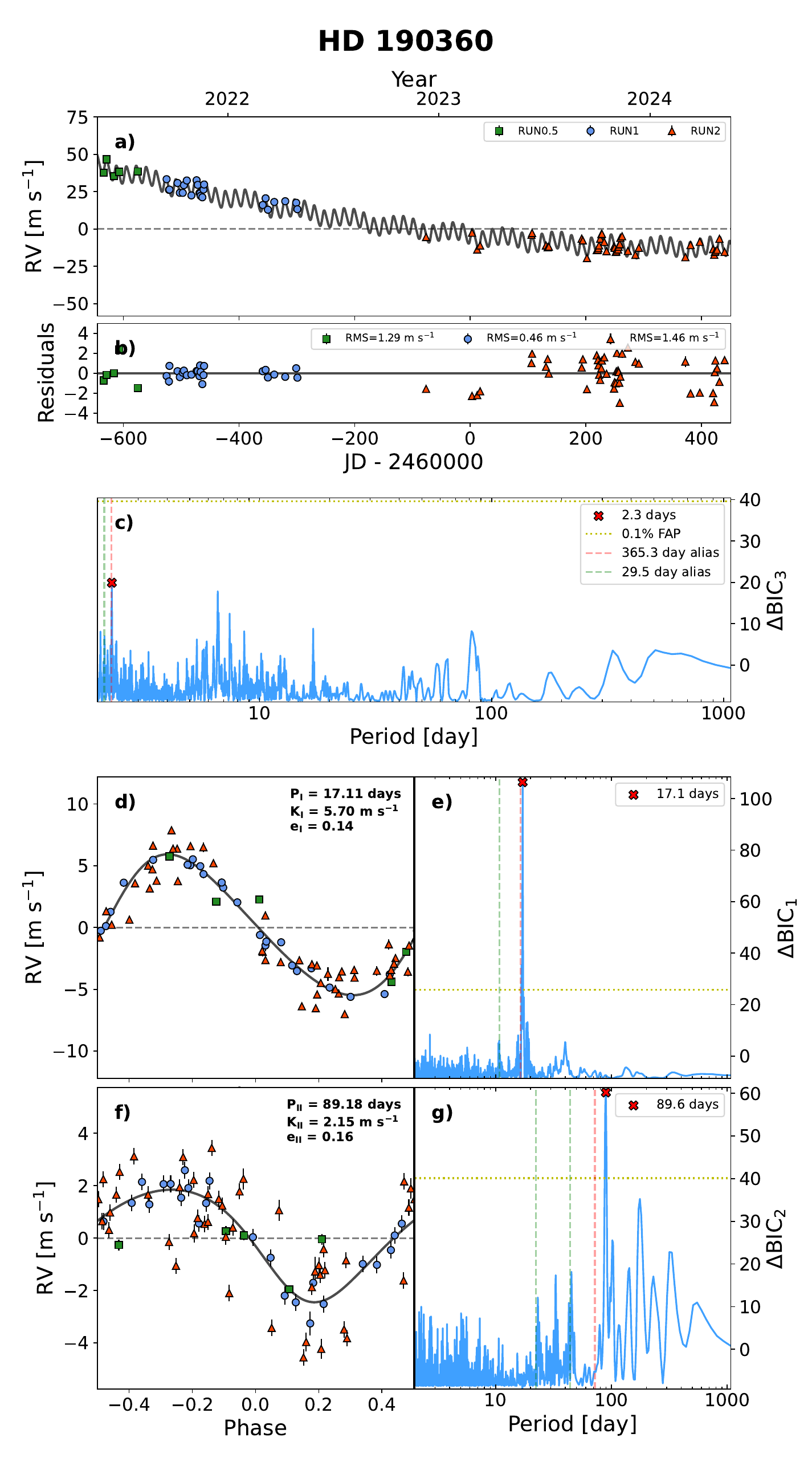}
    \caption{\texttt{RVSearch} results for HD\,190360 for the case in which the long-period signal from the known planet HD\,190360\, was accounted for via a fitted trend. See the caption of \autoref{fig:HD143761_rvsearch} for panel descriptions. In addition to the known planet HD\,190360\,c, we detect one additional signal, which is consistent with a new planet candidate identified by \citet{Giovinazzi2025}. The complete figure set includes results for the blind \texttt{RVSearch} run for all stars with known signals from long-period companions.}    \label{fig:HD190360_rvsearch_blind}
\end{figure}

\begin{figure}
    \centering
    \includegraphics[width=\linewidth]{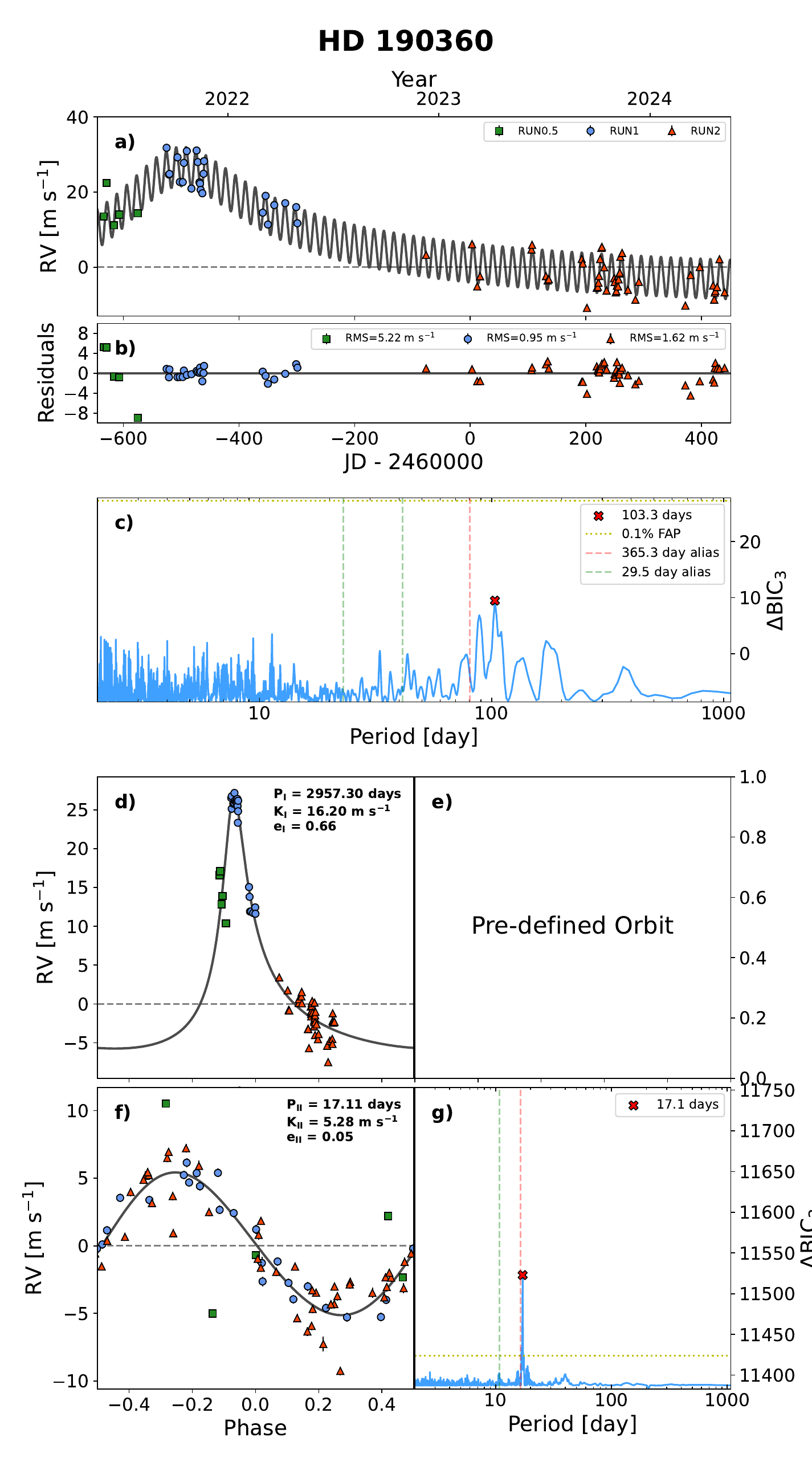}
    \caption{Same as \autoref{fig:HD190360_rvsearch_blind}, but for the case in which a pre-defined orbit for the long-period planet HD\,190360\,b was used in place of a goodness-of-fit trend test. Panels are laid out as described in the caption of \autoref{fig:HD143761_rvsearch}, with the exception of panel d, which now shows the orbit of HD\,190360\,b based on the solution provided by \citet{Giovinazzi2025} We detect no new signals aside from the known planet HD\,190360\,c. The complete figure set includes results for the informed \texttt{RVSearch} run for all stars with known signals from long-period companions.}      \label{fig:HD190360_rvsearch_informed}
\end{figure}

In the informed search, only the 17-day signal of HD\,190360\,c is identified. While a 100-day signal that may be consistent with the candidate is present in the residuals, it does not exceed the false alarm threshold. We show the results of the blind search and informed search in Figures \ref{fig:HD190360_rvsearch_blind} and \ref{fig:HD190360_rvsearch_informed}, respectively. We consider HD\,190360\,c to be successfully detected. 

\input{FigSet9}

\input{FigSet10}

\subsubsection{HD 217107}
HD\,217107 is a G-dwarf hosting two planets, HD\,217107\,b and c, originally discovered by \citet{Fischer1999} and \citet{Vogt2005}, respectively. As with the HD\,190360 system, the outer Jupiter analog (HD\,217107\,c; $P=5139$ days) has a dynamical mass measurement based on a combined RV and astrometric data set \citep{Giovinazzi2025}, but the orbit of the inner planet (HD\,217107\,b; $P=7$ days) does not induce a measurable astrometric signal.

The blind search test prefers a model with no trend/curvature.
Three signals are identified, including a 7-day Keplerian consistent with HD\,217107\,b. The additional signals have best-fit periods of 370 and 913 days, neither of which have any known comparable signals in the literature. However, as with HD\,186427, the RV zero point offset between Run 1 and Run 2 is significantly larger than expected (46 m~s$^{-1}$), and only one signal, corresponding to HD\,217107\,b, is detected in the informed search. The long-period signals identified by the blind search are therefore likely to have resulted from poor fits to the orbit of HD\,217107\,c.
We consider HD\,217107\,b a successful detection.

\subsection{New Candidate Signals}

\subsubsection{HD 10700}
HD\,10700, better known as $\tau$ Ceti, has a contentious history of RV-detected planet candidates. Most recently, \citet{Feng2017} reported four planet candidates with sub-m~s$^{-1}$ semi-amplitudes at orbital periods of 20 ($K=0.49$ m~s$^{-1}$), 49 ($K=0.39$ m~s$^{-1}$), 163 ($K=0.55$ m~s$^{-1}$), and 636 ($K=0.35$ m~s$^{-1}$) days.
The goodness-of-fit test prefers a linear trend of $-0.7$ m~s$^{-1}$~yr$^{-1}$, and in both search cases we detect no significant signals at these periods.
However, both searches do identify a signal at 123 days with a semi-amplitude of 1.3 m~s$^{-1}$ and eccentricity $e=0.7$. \rev{$\tau$ Ceti was recently shown to have a low stellar inclination, such that we are observing the star nearly pole-on relative to the rotation axis \citep{Korolik2023}. At this viewing angle it is unlikely for rotational modulation to affect the RVs at the $>1$ m~s$^{-1}$ level, but the 123-day signal could be an alias of the $46\pm4$-day rotation period as spots move in and out of view on the stellar limb.} We do not see a peak at either of these periods in the GLS periodogram of any of our activity indicators.

\subsubsection{HD 10780}
HD\,10780 straddles the late-G/early-K spectral type boundary and is similar to the Sun in metallicity and log~$g$. A long-term linear trend of $-2.7$ m~s$^{-1}$~yr$^{-1}$ is preferred when we perform the trend test. In both searches, with and without the trend, a signal is identified at 25.6 days and with semi-amplitude $K\sim2$ m~s$^{-1}$, though the simpler model -- without a trend -- converges on a more eccentric orbit.
There are no significant signals near this period in the NEID activity indicators, though there is a hint of a 21-day periodicity in S$_{\rm HK}$.

\subsubsection{HD 4628}
HD\,4628 is an early K-dwarf with a 38.5-day rotation period \citep{Brandenburg2017}. The goodness-of-fit test prefers a linear trend term of $-1.7$ m~s$^{-1}$~yr$^{-1}$. In this case, a 14.7-day signal is identified with a semi-amplitude of 1.14 m/s and an eccentricity of 0.38. In the fit that does not include a trend/curvature term, the strongest peak in the periodogram is at 14.8 days but it does not exceed the false-alarm threshold and is not formally detected.
There are no corresponding periodogram peaks for any of the NEID activity indicators, but we do observe a linear correlation between the RV and S$_{\rm HK}$ measurements which can be attributed to rotational modulation and the long-term stellar activity cycle.

\subsubsection{HD 50692}
For HD\,50692, the trend test prefers a linear model with a long-term trend of $-4.5$ m~s$^{-1}$~yr$^{-1}$. In this case, a super-eccentric ($e=0.96$), 301-day signal was identified with a 10 m~s$^{-1}$ semi-amplitude. In the search with no trend, this signal was not identified, nor is there a peak at this periodicity in the $\Delta$BIC periodogram. Given the inconsistency between these two tests, the validity of this signal is questionable. 

\subsubsection{HD 55575}
For the star HD\,55575, our initial test prefers a model without a trend. In both searches, we identify a 52.5-day signal with $K=1.2$ m~s$^{-1}$ and $e=0.1$. \citet{Rosenthal2021} report the detection of a RV signal at this period which they attribute to the stellar rotation period. However, there is no published rotation period for this star, and neither the HIRES S-index measurements presented by \citet{Rosenthal2021} nor any of the NEID activity indicators presented herein exhibit any periodicity at 52 days or at any aliases thereof. A more thorough investigation is warranted to re-assess the classification of this signal.

\subsubsection{HD 221354}
Our goodness-of-fit trend test returns a linear trend term of $-1.0$ m~s$^{-1}$~yr$^{-1}$, and both searches detect a 1 m~s$^{-1}$, $e=0.5$ signal at 91 days. This signal is likely an alias of the lunar cycle and the 42-day rotation period, which we detect in both the S$_{\rm HK}$ and H$\alpha$ NEID activity indicators.

\subsubsection{HD 146233}
\begin{figure}
    \centering
    \includegraphics[width=\linewidth]{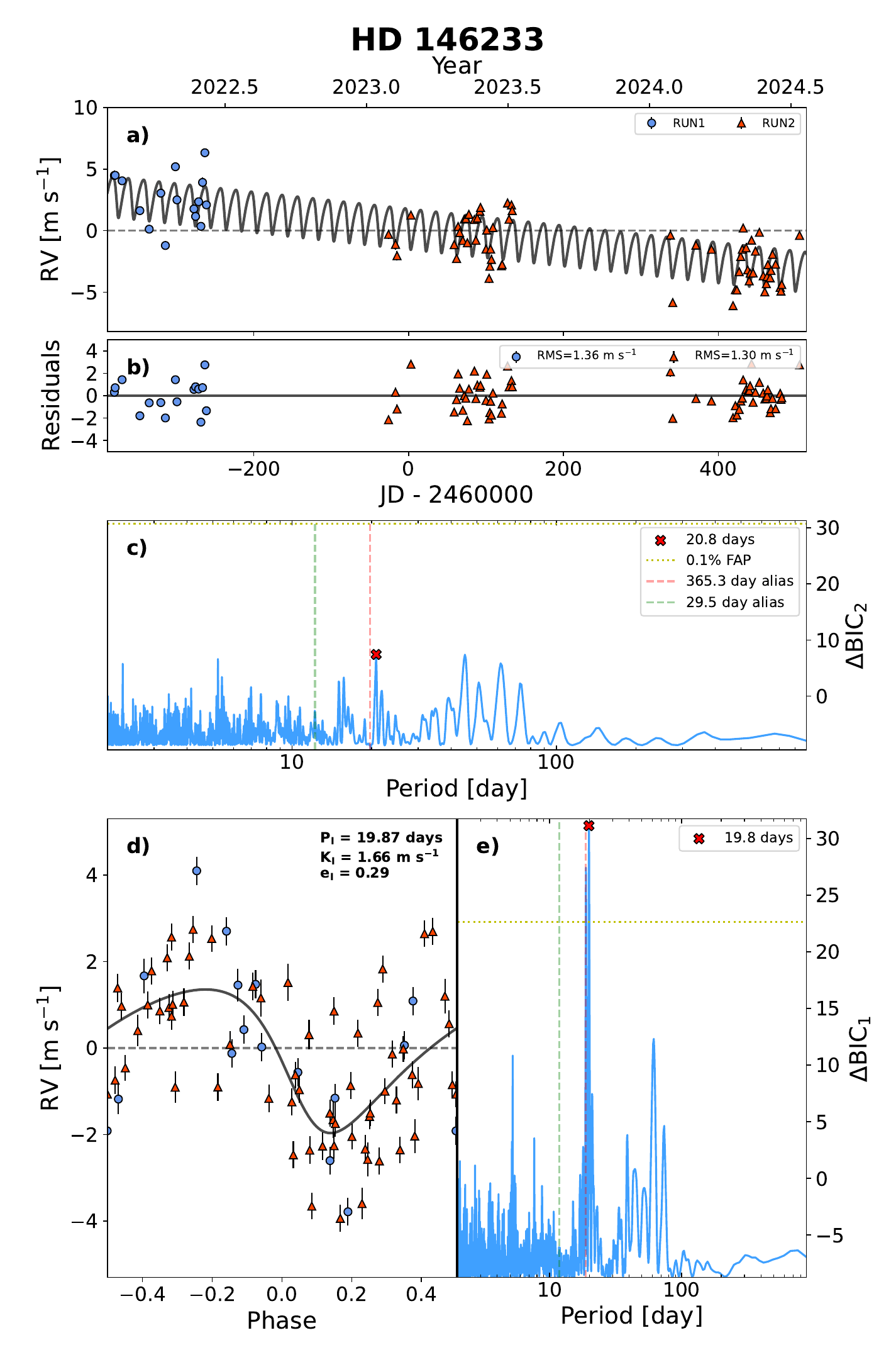}
    \caption{\texttt{RVSearch} results for HD\,146233 for the case in which a goodness-of-fit trend test was allowed and the period search was restricted to $P<\tau$. See the caption of \autoref{fig:HD143761_rvsearch} for panel descriptions. A long-term linear trend was identified and we detect a single planet at $P=19.9$ days, consistent with the candidate reported by \citet{Laliotis2023}.  The complete figure set includes results for the \texttt{RVSearch} run with a trend test for all stars with no known signals from long-period companions.}
    \label{fig:HD146233_rvsearch}
\end{figure}

Using RV measurements of HD\,146233 from HARPS-N, HIRES, PFS, and UCLES in a  blind search analysis similar to the one implemented here, \citet{Laliotis2023} identified a candidate planetary signal with $P=19.9$ days, $K=1.73$ m~s$^{-1}$, and $e=0.38$. For our search that allows for a trend, we detect this same 19.9-day signal ($K=1.6$ m~s$^{-1}$, $e=0.3$) alongside a long-term linear trend of $-2.5$ m~s$^{-1}$~yr$^{-1}$.  The search results are shown in \autoref{fig:HD146233_rvsearch}. The sign and slope of the long-term trend are consistent with the known long-term activity cycle for HD\,146233 as well as a moderate correlation between the NEID RVs and S$_{\rm HK}$ measurements (Spearman's $\rho=0.65$ (Run 1) and $\rho=0.66$ (Run 2)). In our search without a trend, we do observe a peak in the $\Delta$BIC periodogram at 19.9 days, but this peak does not exceed the false alarm threshold.

\input{FigSet11}

\subsubsection{HD 110897}
The goodness-of-fit test for HD\,110897 prefers a trend model 
with a linear term of $-2.3$ m~s$^{-1}$~yr$^{-1}$ and a quadratic term of $-0.2$ m~s$^{-1}$~yr$^{-2}$. A 358-day signal is detected with $K=2.5$ m~s$^{-1}$ and $e=0.69$. For the second search, with no long-term trend, A 502-day signal is detected with $K=3.7$ m~s$^{-1}$ and $e=0.57$. In both cases, following the removal of the detected long-period signal, the next most prominent peak is at $P\sim36$ days, which coincides with a significant S$_{\rm HK}$ periodicity for this star. We also observe a moderate linear correlation between the RVs and S$_{\rm HK}$ values (Spearman's $\rho=0.68$ (Run 1) and $\rho=0.73$ (Run 2)), indicating that much of the observed RV variation is driven by stellar activity. A more detailed analysis informed by both the RV and activity index measurements is necessary to discern the true nature of any long-period signals.

\subsubsection{HD 126053}

HD\,126053 is a Solar-type star with a 26-day rotation period \citep{Hempelmann2016}. Our initial test prefers a model with no trend; in both searches, we detect a 10.9-day RV signal with $K=1.1$ m~s$^{-1}$ and $e=0.2$. While we do see power near the stellar rotation period in multiple activity indicators, including S$_{\rm HK}$ and H$\alpha$, the rotation signal is absent from the RVs.

\subsubsection{HD 9407}

Our initial search prefers a model with a linear trend of $-1.0$ m~s$^{-1}$~yr$^{-1}$ for HD\,9407. This test returns a detection of a 61-day RV signal with $K=1.0$ m~s$^{-1}$ and $e=0.55$, while the search with no trend returns
a 52-day RV signal with $K=1.1$ m~s$^{-1}$ and $e=0.51$. No comparable signals are seen in any of the activity indicator metrics, and the activity and RVs appear uncorrelated.

\subsubsection{HD 127334}
The trend test prefers a second-order model with a linear term of $0.7$ m~s$^{-1}$~yr$^{-1}$ and quadratic term of $-0.7$ m~s$^{-1}$~yr$^{-2}$. For this search, a 185-day signal is identified with a $K=0.6$ m~s$^{-1}$ and $e=0.3$. In the no trend search, a signal is also identified but at 371 days with a similar semi-amplitude and eccentricity ($K=0.6$ m~s$^{-1}$, $e=0.5$). However, we note that these signals are aliases of each other, where power is transferred from one period to the other due to the inclusion (or exclusion) of the quadratic trend; a significant 184-day peak is also present in the $\Delta$BIC periodogram for the latter search, and the power at this period is removed when we subtract the 371-day signal. Further, significant long-period power is present for several of the NEID activity indicators, including both S$_{\rm HK}$ and H$\alpha$, suggesting that the observed RV variation is driven by stellar activity and not a planetary companion.

\subsubsection{HD 185144}

HD\,185144 ($\sigma$ Dra) is an early K-dwarf with no known planetary companions, though the star does have a well-characterized, $\sim6$-yr stellar activity cycle \citep{Baum2022,Isaacson2024}. The search with a trend adopts a quadratic model with linear and quadratic trend terms of $1.11$ m~s$^{-1}$~yr$^{-1}$ and $-0.68$ m~s$^{-1}$~yr$^{-2}$, and no signals are detected. Without a trend, the search returns a significant signal at $P=409.7$ d, with orbital eccentricity $e=0.75$ and semi-amplitude $K=1.34$ m~s$^{-1}$. Significant long-period power is present in the stellar activity indicators, corresponding to a change in the overall activity level as the star enters an activity minimum. The quadratic trend and the detected eccentric signal could be explained by the RV response to the stellar activity cycle, but further investigation is warranted.

\subsubsection{HD 68017}
HD\,68017 is a G-dwarf with an M-dwarf companion on a very close orbit \citep[0.59'', projected separation 13 au;][]{Crepp2012}. In both search cases ("blind" with best-fit linear and quadratic trend terms of $15.7$ m~s$^{-1}$~yr$^{-1}$ and $-3.63$ m~s$^{-1}$~yr$^{-2}$, and "informed" of the known companion parameters) a 118-day signal is identified with $K = 1.4$ m~s$^{-1}$, $e = 0.6$ and $K = 1.2$ m~s$^{-1}$, $e = 0.5$ for the blind and informed saerches, respectively.
We do not see peaks at this period for any of the activity indicators, nor are the S-index and RVs linearly correlated.  In addition, a signal at this period is also present in the RV residuals to the \citet{Giovinazzi2025} joint RV+astrometric stellar orbit fit, although this signal does not exceed the false alarm threshold. This signal merits further investigation to determine whether it is of planetary origin.

\subsection{Known Activity Signals}\label{sec:known_activity}

We describe below any signals that were identified in our search and that were already known to be linked to the stellar rotation period or activity cycle.

\subsubsection{HD 26965}
Multiple studies have investigated the nature of a $\sim$42-day signal appearing in the RVs of HD\,26965. While this was initially reported as a candidate super-Earth, HD\,26965\,b by \citet{Diaz2018} and subsequently confirmed by \citet{Ma2018}, other studies have posited that the RVs are driven by stellar activity \citep{Zhao2022,Laliotis2023}. Most recently, multiple independent analyses of spectrum-level variations in the Run 1 NEID data have firmly classified this as a rotationally-modulated activity signal \citep{Burrows2024,Siegel2024,Gilbertson2024}.

When we allow for a trend, a linear model with a slope of $-1.8$ m~s$^{-1}$~yr$^{-1}$ was preferred. The 42-day rotation signal is identified in both searches, and no additional significant signals are present in the residuals.

\subsubsection{HD 38858}
\citet{Mayor2011} reported the detection of a super-Earth around HD\,38858 on an eccentric ($e=0.27$) orbit with a period of 407 days, but subsequent analyses show that this was an alias of the stellar activity cycle \citep{Kennedy2015,Flores2018}.

Our trend test prefers a second order model with a linear term of $1.26$ m~s$^{-1}$~yr$^{-1}$ and a quadratic term of $-1.40$ m~s$^{-1}$~yr$^{-2}$. In our searches with and without the trend, 415-day and 423-day signals are identified, respectively, with different orbital eccentricities and semi-amplitudes. There is a significant peak at this period in the periodograms of multiple NEID activity indicators, further supporting the non-planetary nature of this signal. 

\begin{deluxetable*}{lrrrrrrrrr}
\tabletypesize{\scriptsize}
\label{tab:rvsearch_results_trend}
\tablecaption{\texttt{RVSearch} Detected Signals (trend allowed)}
\tablehead{
\colhead{Name} & \colhead{$P$ (d)} &\colhead{$e$} & \colhead{$\omega$ ($^{\circ}$)} & \colhead{$K$ (m\,s$^{-1}$)}  & \colhead{$T_{\rm conj, inf}$ (d)} & \colhead{$\Delta\gamma_{\rm{Run1-Run0.5}}$ (m\,s$^{-1}$)} & \colhead{$\Delta\gamma_{\rm{Run1-Run2}}$ (m\,s$^{-1}$)} & 
\colhead{$\dot{\gamma}$ (m\,s$^{-1}$\,yr$^{-1}$)} & \colhead{$\ddot{\gamma}$ (m\,s$^{-1}$\,yr$^{-2}$)}
}
\startdata
HD 4628 I & $14.7$ & $0.38$ & $-16.5$ & $1.14$ & $2460051.362$ & \nodata & $0.34$ & $-1.68$ & \nodata \\
HD 9407 I & $61.4$ & $0.55$ & $50.5$ & $0.97$ & $2459996.387$ & $2.85$ & $-0.23$ & $-1.00$ & \nodata \\
HD 10700 I & $123.3$ & $0.71$ & $-167.0$ & $1.25$ & $2460243.447$ & $0.60$ & $0.23$ & $-0.69$ & \nodata \\
HD 10780 I & $25.6$ & $0.14$ & $127.0$ & $2.13$ & $2459962.367$ & \nodata & $-5.04$ & $-2.74$ & \nodata \\
HD 26965 I & $41.6$ & $0.29$ & $-173.0$ & $2.41$ & $2459955.326$ & \nodata & $-0.26$ & $-1.80$ & \nodata \\
HD 38858 I & $415.7$ & $0.09$ & $48.0$ & $3.33$ & $2460145.907$ & \nodata & $4.27$ & $1.34$ & $-1.41$ \\
HD 50692 I & $301.3$ & $0.96$ & $128.0$ & $10.00$ & $2459835.477$ & \nodata & $-2.95$ & $-4.52$ & \nodata \\
HD 55575 I & $52.5$ & $0.08$ & $-94.8$ & $1.18$ & $2459977.258$ & \nodata & $1.04$ & \nodata & \nodata \\
HD 68017 I & $118.1$ & $0.58$ & $142.0$ & $1.39$ & $2460110.162$ & \nodata & $2.56$ & $15.70$ & $-3.61$ \\
HD 86728 I & $31.2$ & $0.22$ & $109.0$ & $2.13$ & $2460040.198$ & $1.22$ & $1.25$ & \nodata & \nodata \\
HD 95735 I & $13.0$ & $0.09$ & $-67.0$ & $1.42$ & $2460045.233$ & $4.48$ & $2.65$ & \nodata & \nodata \\
HD 95735 II & $654.0$ & $0.53$ & $-162.0$ & $2.48$ & $2460204.195$ & $4.48$ & $2.65$ & \nodata & \nodata \\
HD 95735 III & $219.9$ & $0.17$ & $78.0$ & $1.24$ & $2460097.917$ & $4.48$ & $2.65$ & \nodata & \nodata \\
HD 110897 I & $357.4$ & $0.69$ & $160.0$ & $2.51$ & $2460266.556$ & \nodata & $-2.54$ & $-2.30$ & $-0.16$ \\
HD 115617 I & $38.0$ & $0.07$ & $169.0$ & $3.78$ & $2460064.264$ & $3.87$ & $0.77$ & \nodata & \nodata \\
HD 115617 II & $4.2$ & $0.10$ & $85.6$ & $2.49$ & $2460056.265$ & $3.87$ & $0.77$ & \nodata & \nodata \\
HD 115617 III & $92.8$ & $0.30$ & $120.0$ & $1.78$ & $2460042.386$ & $3.87$ & $0.77$ & \nodata & \nodata \\
HD 115617 IV & $395.0$ & $0.44$ & $178.0$ & $4.32$ & $2460230.544$ & $3.87$ & $0.77$ & \nodata & \nodata \\
HD 126053 I & $10.9$ & $0.20$ & $26.8$ & $1.13$ & $2460076.299$ & \nodata & $0.95$ & \nodata & \nodata \\
HD 127334 I & $185.3$ & $0.29$ & $26.2$ & $0.58$ & $2460117.623$ & $-0.46$ & $1.87$ & $0.66$ & $-0.67$ \\
HD 143761 I & $39.9$ & $0.04$ & $-82.4$ & $67.40$ & $2460062.068$ & $0.26$ & $1.02$ & \nodata & \nodata \\
HD 143761 II & $102.9$ & $0.09$ & $6.1$ & $3.60$ & $2460025.102$ & $0.26$ & $1.02$ & \nodata & \nodata \\
HD 143761 III & $12.9$ & $0.63$ & $69.4$ & $1.57$ & $2460069.576$ & $0.26$ & $1.02$ & \nodata & \nodata \\
HD 146233 I & $19.9$ & $0.29$ & $130.0$ & $1.66$ & $2460099.309$ & \nodata & $0.88$ & $-2.50$ & \nodata \\
HD 168009 I & $15.1$ & $0.15$ & $-92.8$ & $2.25$ & $2460076.966$ & \nodata & $-2.83$ & $-3.03$ & \nodata \\
HD 168009 II & $61.3$ & $0.57$ & $177.0$ & $2.62$ & $2460006.727$ & \nodata & $-2.83$ & $-3.03$ & \nodata \\
HD 186427 I & $370.3$ & $0.69$ & $110.0$ & $67.40$ & $2460131.739$ & \nodata & $-124.00$ & $-66$ & $0.93$ \\
HD 190360 I & $17.1$ & $0.14$ & $-73.8$ & $5.70$ & $2460128.541$ & $6.41$ & $-6.71$ & $-18.30$ & $6.57$ \\
HD 190360 II & $89.2$ & $0.16$ & $151.0$ & $2.15$ & $2460087.656$ & $6.41$ & $-6.71$ & $-18.30$ & $6.57$ \\
HD 217107 I & $7.1$ & $0.13$ & $23.7$ & $142.00$ & $2460190.993$ & \nodata & $46.10$ & \nodata & \nodata \\
HD 217107 II & $370.9$ & $0.68$ & $-147.0$ & $10.80$ & $2460462.273$ & \nodata & $46.10$ & \nodata & \nodata \\
HD 217107 III & $913.2$ & $0.26$ & $-41.1$ & $14.30$ & $2460143.295$ & \nodata & $46.10$ & \nodata & \nodata \\
HD 221354 I & $91.5$ & $0.47$ & $-148.0$ & $1.04$ & $2460098.621$ & $1.16$ & $-0.34$ & $-1.04$ & \nodata \\
\enddata
\tablenotetext{}{\textbf{Note.} For each star, signals are listed in order of detection with \texttt{RVSearch}. Columns are orbital period ($P$), eccentricity ($e$), argument of periastron ($\omega$), RV semi-amplitude ($K$), time of inferior conjunction ($T_{\rm conj,inf}$), systemic RV offsets between NEID RV eras ($\Delta\gamma$), and best-fit linear ($\dot{\gamma}$) and quadratic ($\ddot{\gamma}$) trend terms.}\end{deluxetable*}

\begin{deluxetable*}{lrrrrrrrr}
\tabletypesize{\scriptsize}
\label{tab:rvsearch_results_no_trend}
\tablecaption{\texttt{RVSearch} Detected Signals (no trend allowed)}
\tablehead{
\colhead{Name} & \colhead{$P$ (d)} &\colhead{$e$} & \colhead{$\omega$ ($^{\circ}$)} & \colhead{$K$ (m\,s$^{-1}$)} & \colhead{$T_{\rm conj, inf}$ (d)} & \colhead{$\Delta\gamma_{\rm{Run1-Run0.5}}$ (m\,s$^{-1}$)} & \colhead{$\Delta\gamma_{\rm{Run1-Run2}}$ (m\,s$^{-1}$)}
}
\startdata
HD 9407 I & $52.4$ & $0.51$ & $1.4$ & $1.05$ & $2459977.249$ & $0.59$ & $1.72$ \\
HD 10700 I & $123.2$ & $0.70$ & $-167.0$ & $1.30$ & $2460243.055$ & $0.05$ & $1.47$ \\
HD 10780 I & $25.6$ & $0.25$ & $168.0$ & $2.26$ & $2459963.705$ & \nodata & $-0.49$ \\
HD 26965 I & $41.6$ & $0.25$ & $-168.0$ & $2.53$ & $2459954.587$ & \nodata & $2.59$ \\
HD 38858 I & $423.3$ & $0.16$ & $45.5$ & $3.86$ & $2460149.433$ & \nodata & $1.11$ \\
HD 55575 I & $52.5$ & $0.08$ & $-94.4$ & $1.18$ & $2459977.246$ & \nodata & $1.04$ \\
HD 68017 I & $117.7$ & $0.49$ & $142.0$ & $1.23$ & $2460107.517$ & \nodata & $2.35$ \\
HD 86728 I & $31.2$ & $0.22$ & $110.0$ & $2.13$ & $2460040.230$ & $1.22$ & $1.25$ \\
HD 95735 I & $13.0$ & $0.14$ & $-38.0$ & $1.40$ & $2460045.014$ & $1.92$ & $-2.64$ \\
HD 110897 I & $502.4$ & $0.57$ & $90.9$ & $3.71$ & $2460252.240$ & \nodata & $-2.04$ \\
HD 115617 I & $38.0$ & $0.07$ & $172.0$ & $3.75$ & $2460064.204$ & $3.85$ & $0.87$ \\
HD 115617 II & $4.2$ & $0.11$ & $91.7$ & $2.50$ & $2460056.273$ & $3.85$ & $0.87$ \\
HD 115617 III & $92.7$ & $0.28$ & $120.0$ & $1.76$ & $2460041.959$ & $3.85$ & $0.87$ \\
HD 115617 IV & $390.2$ & $0.44$ & $177.0$ & $5.62$ & $2460217.207$ & $3.85$ & $0.87$ \\
HD 126053 I & $10.9$ & $0.21$ & $26.3$ & $1.13$ & $2460076.287$ & \nodata & $0.95$ \\
HD 127334 I & $371.4$ & $0.49$ & $-86.2$ & $0.57$ & $2460219.897$ & $1.44$ & $1.50$ \\
HD 143761 I & $39.9$ & $0.04$ & $-82.4$ & $67.40$ & $2460062.070$ & $0.25$ & $1.02$ \\
HD 143761 II & $102.9$ & $0.09$ & $5.3$ & $3.59$ & $2460025.057$ & $0.25$ & $1.02$ \\
HD 143761 III & $12.9$ & $0.63$ & $69.5$ & $1.57$ & $2460069.576$ & $0.25$ & $1.02$ \\
HD 185144 I & $409.7$ & $0.75$ & $166.0$ & $1.34$ & $2460294.541$ & $2.27$ & $2.05$ \\
HD 190360 I & $17.1$ & $0.05$ & $-56.2$ & $5.28$ & $2460128.774$ & $-16.20$ & $3.49$ \\
HD 217107 I & $7.1$ & $0.13$ & $23.5$ & $142.0$ & $2460190.995$ & \nodata & $1.73$ \\
HD 221354 I & $91.5$ & $0.50$ & $-148.0$ & $1.12$ & $2460007.251$ & $0.76$ & $1.49$ \\
\enddata
\tablenotetext{}{\textbf{Note.} For each star, signals are listed in order of detection with \texttt{RVSearch}. Columns are the same as in Table \ref{tab:rvsearch_results_trend}, with the exception of the trend terms which are not included here.}
\end{deluxetable*}

\section{Discussion}\label{sec:discussion}

\subsection{Sensitivity to known exoplanets}

\begin{figure*}
    \centering
    \includegraphics[width=\linewidth]{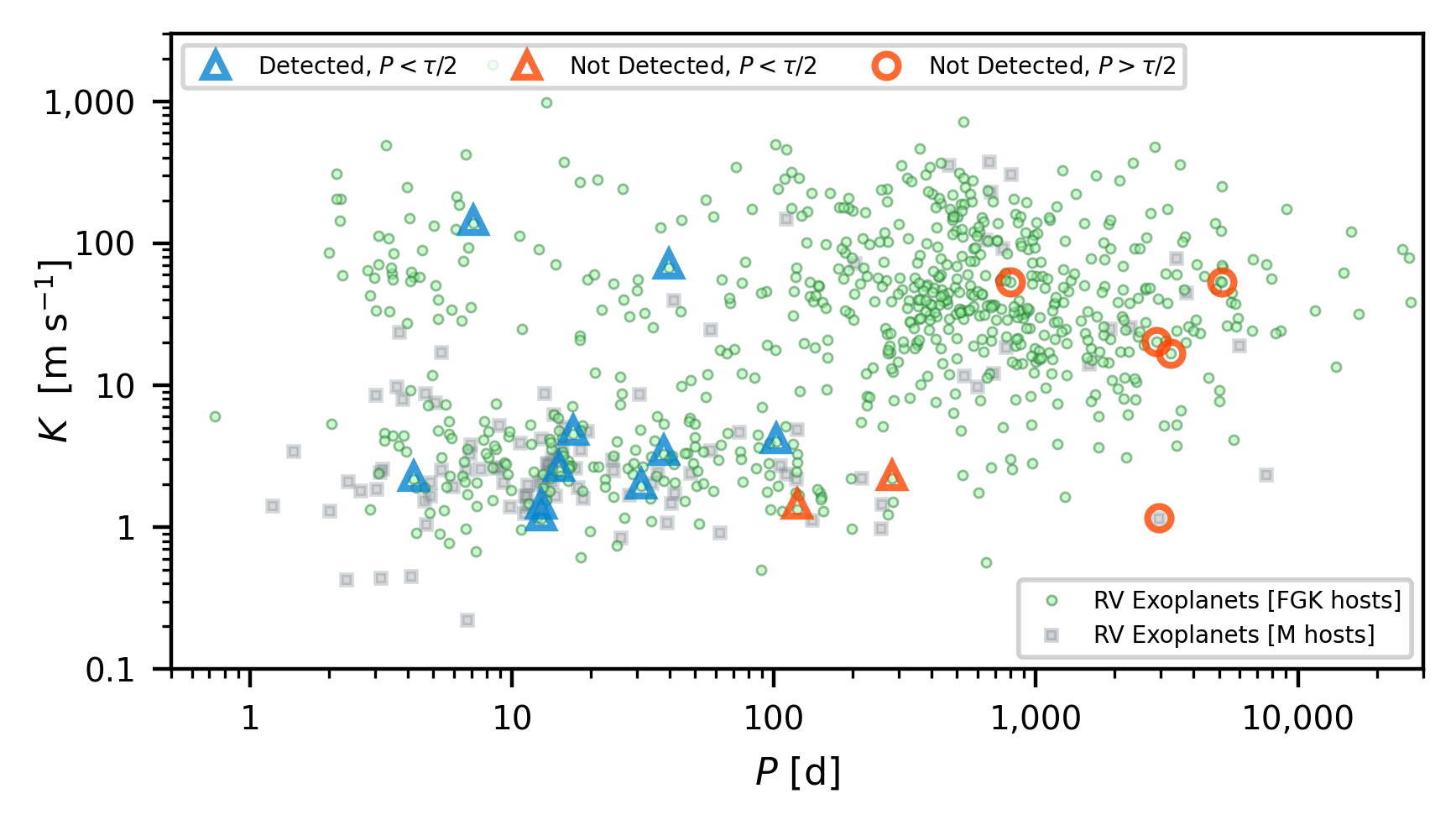}
    \caption{Orbital periods and semi-amplitudes for all known exoplanets discovered via the RV method. Planetary companions to FGK host stars are shown as green circles and companions to M-dwarfs are shown as gray squares. Planets orbiting to NETS targets stars that are successfully detected are highlighted with blue triangles (all with $P<\tau/2$), and planets that we fail to independently recover are marked with orange triangles ($P<\tau/2$) and orange circles ($P>\tau/2$). HD 143761\,d \rev{and HD\,115617\,d are} the only planets with $P<\tau/2$ that we fail to detect. These results suggest that the present survey sensitivity is limited by the observing baseline, not by the achieved RV precision.}
    \label{fig:detection_summary}
\end{figure*}

Our analysis in Sections \ref{sec:rvsearch} and \ref{sec:known_signals} demonstrates that with just 3 years of dedicated NEID observations, our survey is approaching the same sensitivity as all cumulative RV monitoring of these same target stars over the past few decades. We successfully detect 10 of the \rev{12} short-period ($P\leq\tau/2$) planets (period less than half the NEID baseline) in our sample (\autoref{fig:detection_summary}), even in cases where no prior information is provided about long-period signals.
In addition, we note that all of these detections, including several with $K\lesssim2$ m~s$^{-1}$, were achieved without any careful treatment (i.e., modeling, mitigation) of stellar variability. This is a testament to the advantages of selecting quiet, inactive target stars for the survey and observing with a highly stabilized instrument.

None of the five long-period planets (period greater than half the NEID baseline) are recovered (\autoref{fig:detection_summary}), but this is unsurprising given that in most cases NETS has only covered a small fraction of their orbits. This will continue to be true through the end of the five-year survey, where only HD\,186427\,b ($P=799$ d) will end up with full coverage over multiple orbits. Still, as \citet{Giovinazzi2025} show, NEID data can be combined with archival data over longer baselines to place stringent constraints on the orbits of long-period planets. Further, \autoref{fig:detection_summary}  illustrates that the survey sensitivity is presently limited by the observing baseline, not by the RV precision or the intrinsic variability of the target stars. As the survey progresses, we expect to be able to robustly detect planets with $K=1-2$ m~s$^{-1}$ out to periods of several hundred days. 

\rev{\subsection{New Candidate Signals}}

\rev{We report new candidate RV signals alongside known signals in Tables \ref{tab:rvsearch_results_no_trend} and \ref{tab:rvsearch_results_trend}. These new signals are described briefly in Section \ref{sec:known_signals}, but we refrain from classifying them as either real planets or false positives and we defer detailed analysis to future studies. However, it is worth noting that unlike the known planet signals we recover, which generally have eccentricities consistent with the literature values, many of the new signals we report have fairly eccentric orbits. This is atypical for the known exoplanet population, and may be an outcome of our use of a flat eccentricity prior, which is known to allow for spurious eccentric orbits when the periastron phase is poorly sampled.  Any follow-up investigation of these signals should include an analysis of whether the high eccentricities persist when different priors are imposed.}

\subsection{Targets on ExEP List}

The 2020 Decadal Survey of Astronomy and Astrophysics \citep{Decadal2020} recommended the development and construction of the Habitable Worlds Observatory (HWO), a flagship space telescope with high contrast direct imaging and spectroscopy capabilities from UV to IR wavelengths. This telescope will carry out a survey of nearby stars to search for and and characterize the atmospheres of at least 25 habitable-zone, terrestrial exoplanets. HWO is presently expected to be able to accomplish this through an uninformed search. However, information gleaned from EPRV surveys such as NETS has the potential to greatly increase the efficiency and scientific return of the mission. As \citet{Morgan2021} show, partial prior knowledge of Earth-like planet orbits from RV measurements can be used to improve HWO survey yield (i.e., the number of characterized planets) by up to 50\% and accelerate the detection of key atmospheric biomarkers. Further, recent work by \citet{Kane2024} has highlighted the obstacles posed by as-yet undetected massive planets orbiting in or near the habitable zone. Planets as small as $10\,{\rm M}_\oplus$ can dynamically preclude stable orbits for terrestrial planets through much of the habitable zones of their host stars, rendering these targets unfit for the HWO survey. But for the majority of high priority HWO target stars identified by \citet{Mamajek2024}, existing RV data are insufficient to identify or rule out such perturbers \citep{Laliotis2023,Harada2024}.

Independent of these challenges, accurate interpretation of atmospheric spectra collected by HWO or similar missions will require precise knowledge of the exoplanets' physical properties. In particular, insufficient constraints on the planetary mass can lead to large uncertainties in the inferred atmospheric parameters \citep{Batalha2019} and incorrect assessments of habitability and of the presence or absence of biosignatures \citep{Damiano2025}. \citet{Damiano2025} recommend a mass precision of better than 10\% to guarantee reliable results from reflected light spectroscopy of Earth-like planets.

The NETS target list includes 23 high-priority HWO target stars, including many with known planets (HD\,115617, HD\,143761, HD\,190360, HD\,86728, HD\,95735) and others with promising, low-amplitude candidate RV signals identified in this work and previous works (HD\,4628, HD\,10780, HD\,146233). It is essential that these stars continue to be monitored with NETS and other EPRV surveys to guarantee that we have an accurate census of their planetary systems and that these systems are well characterized in preparation for HWO. At the very least, establishing and maintaining an EPRV baseline now will greatly accelerate mass measurements for Earth-like planets independently detected by HWO. NETS data can also serve an important role in continuing to study the impact of stellar variability on the spectra of these host stars; while the NETS targets are quiet, they are not silent, and activity-driven spectral variations are well known to complicate exoplanet spectroscopy measurements \citep{Rackham2018,Rackham2019}.

\section{Summary and Conclusions}

In this work, we have detailed the progress and performance of the NEID Earth Twin Survey through its first three years on sky, and we present the full set of RVs and accompanying activity indicators measurements for our carefully curated sample of bright, RV-quiet stars. We outline the motivation behind the NETS observing strategies and we describe how these observations have been implemented within the constraints of the NEID queue. Although the execution rate was heavily impacted by an extended facility closure, survey progress prior to and following this interruption has been promising.

We show that NEID achieves sub-m~s$^{-1}$ RMS RV scatter for 10 stars, reaching values as low as 60 cm~s$^{-1}$ over baselines of $\tau>1$ yr.
Our \texttt{RVSearch} analysis demonstrates that the NETS data set alone can blindly recover nearly all known planets with orbits shorter than the time baseline of the survey, including multiple with signals weaker than $K=2$ m~s$^{-1}$.
These results are achieved using measurements taken directly from the NEID DRP, with no post-processing, and they are a testament to the exceptional performance of the spectrograph, survey design, and pipeline. As anticipated, we do not independently recover long-period planets, where the orbit exceeds half the existing NEID baseline. Still, as evidenced by the RMS of the RV residuals to the published orbit models, it is clear that the NETS RVs are in good agreement with these long-period signals. 

In addition, we present a chromatic analysis of the RV zero point offset introduced following the Contreras wildfire in 2022, and we show that for FGK-dwarfs, NEID is highly resilient to disruptions of this magnitude. We also describe the discovery and characterization of an earlier RV zero point offset in 2021, which highlights the utility of large data sets like NETS for characterizing the spectrograph itself. Our constraints on these offsets will also inform any future analyses using RVs that span multiple NEID eras, mitigating degeneracies between the offsets and long-period signals.

We also identify a number of RV signals that have not yet been discussed in the literature, some of which seem likely to correspond to new exoplanets. Further investigation of these signals is warranted, and our team will address each of them in detail in future publications. Future, tailored analyses of individual systems will incorporate archival RV measurements and newly acquired NETS RVs as well as considerations for stellar activity, using the indicators provided as part of the NETS data set where appropriate.

\section{Acknowledgments}

This paper contains data taken with the NEID instrument, which was funded by the NASA-NSF Exoplanet Observational Research (NN-EXPLORE) partnership and built by Pennsylvania State University. 
NEID is installed on the WIYN telescope, which is operated by the NSF's National Optical-Infrared Astronomy Research Laboratory, and the
NEID archive is operated by the NASA Exoplanet Science Institute at the California Institute of Technology. NEID is funded by NASA through JPL contract 1547612 and the NEID Data Reduction Pipeline is funded through JPL contract 1644767.
NN-EXPLORE is managed by the Jet Propulsion Laboratory, California Institute of Technology under contract with the National Aeronautics and Space Administration. We thank the NEID Queue Observers and WIYN Observing Associates for their skillful execution of our observations.
Part of this work was performed at the Jet Propulsion Laboratory, California Institute of Technology, sponsored by the United States Government under the Prime Contract 80NM0018D0004 between Caltech and NASA.
CIC acknowledges support by an appointment to the NASA Postdoctoral Program at the Goddard Space Flight Center, administered by ORAU through a contract with NASA.

Computations for this research were performed on the Pennsylvania State University's Institute for Computational and Data Sciences Advanced Cyberinfrastructure (ICDS-ACI). This content is solely the responsibility of the authors and does not necessarily represent the views of the Institute for Computational and Data Sciences. The Center for Exoplanets and Habitable Worlds and the Penn State Extraterrestrial Intelligence Center are supported by the Pennsylvania State University and the Eberly College of Science. This work has made use of data from the European Space Agency (ESA) mission Gaia, processed by the Gaia Data Processing and Analysis Consortium (DPAC). Funding for the DPAC has been provided by national institutions, in particular the institutions participating in the Gaia Multilateral Agreement. This research has made use of the SIMBAD database, operated at CDS, Strasbourg, France, and NASA's Astrophysics Data System Bibliographic Services.

Based in part on observations at Kitt Peak National Observatory, NSF’s NOIRLab, managed by the Association of Universities for Research in Astronomy (AURA) under a cooperative agreement with the National Science Foundation. The authors are honored to be permitted to conduct astronomical research on Iolkam Du’ag (Kitt Peak), a mountain with particular significance to the Tohono O’odham.
We also express our deepest gratitude to Zade Arnold, Joe Davis, Michelle Edwards, John Ehret, Tina Juan, Brian Pisarek, Aaron Rowe, Fred Wortman, the Eastern Area Incident Management Team, and all of the firefighters and air support crew who fought the recent Contreras fire. Against great odds, you saved Kitt Peak National Observatory.

The Pennsylvania State University campuses are located on the original homelands of the Erie, Haudenosaunee (Seneca, Cayuga, Onondaga, Oneida, Mohawk, and Tuscarora), Lenape (Delaware Nation, Delaware Tribe, Stockbridge-Munsee), Shawnee (Absentee, Eastern, and Oklahoma), Susquehannock, and Wahzhazhe (Osage) Nations.  As a land grant institution, we acknowledge and honor the traditional caretakers of these lands and strive to understand and model their responsible stewardship. We also acknowledge the longer history of these lands and our place in that history.

\facilities{WIYN (NEID)}

\software{\texttt{astropy} \citep{AstropyCollaboration2013,AstropyCollaboration2018,AstropyCollaboration2022}, \citep{Kanodia2018}, \texttt{matplotlib} \citep{Hunter2007}, \texttt{numpy} \citep{Harris2020}, \texttt{RadVel} \citep{Fulton2018}, \texttt{RVSearch} \citep{Rosenthal2021}, \texttt{scipy} \citep{Oliphant2007}}

\bibliography{references}{}
\bibliographystyle{aasjournalv7}

\end{document}

%% file: affil_list.tex
\newcommand{\PSUAA}{Department of Astronomy \& Astrophysics, 525 Davey Laboratory, 251 Pollock Road, Penn State, University Park, PA, 16802, USA}
\newcommand{\PSUCEHW}{Center for Exoplanets and Habitable Worlds, 525 Davey Laboratory, 251 Pollock Road, Penn State, University Park, PA, 16802, USA}
\newcommand{\PSUARC}{Astrobiology Research Center, 525 Davey Laboratory, 251 Pollock Road, Penn State, University Park, PA, 16802, USA}
\newcommand{\PSETI}{Penn State Extraterrestrial Intelligence Center, 525 Davey Laboratory, 251 Pollock Road, Penn State, University Park, PA, 16802, USA}
\newcommand{\PSUICDS}{Institute for Computational and Data Sciences, Penn State, University Park, PA, 16802, USA}
\newcommand{\PSUCASt}{Center for Astrostatistics, 525 Davey Laboratory, 251 Pollock Road, Penn State, University Park, PA, 16802, USA}
\newcommand{\UA}{Steward Observatory, University of Arizona, 933 N.\ Cherry Ave, Tucson, AZ 85721, USA}

\newcommand{\Penn}{Department of Physics and Astronomy, University of Pennsylvania, 209 S 33rd St, Philadelphia, PA 19104, USA}
\newcommand{\Caltech}{Department of Astronomy, California Institute of Technology, 1200 E California Blvd, Pasadena, CA 91125, USA}

\newcommand{\GoddardESAL}{Exoplanets and Stellar Astrophysics Laboratory, NASA Goddard Space Flight Center, Greenbelt, MD 20771, USA}

\newcommand{\NOIRLab}{U.S. National Science Foundation National Optical-Infrared Astronomy Research Laboratory, 950 N.\ Cherry Ave., Tucson, AZ 85719, USA}

\newcommand{\Macquarie}{School of Mathematical and Physical Sciences, Macquarie University, Balaclava Road, North Ryde, NSW 2109, Australia}

\newcommand{\JPL}{Jet Propulsion Laboratory, California Institute of Technology, 4800 Oak Grove Drive, Pasadena, California 91109}

\newcommand{\UCI}{Department of Physics \& Astronomy, The University of California, Irvine, Irvine, CA 92697, USA}
\newcommand{\Carleton}{Carleton College, One North College St., Northfield, MN 55057, USA}
\newcommand{\Carnegie}{Carnegie Science Earth and Planets Laboratory, 5241 Broad Branch Road, NW, Washington, DC 20015, USA}

\newcommand{\TIFR}{Department of Astronomy and Astrophysics, Tata Institute of Fundamental Research, Homi Bhabha Road, Colaba, Mumbai 400005, India}

\newcommand{\UAm}{Anton Pannekoek Institute for Astronomy, 904 Science Park, University of Amsterdam, Amsterdam, 1098 XH}

\newcommand{\Amherst}{Department of Physics and Astronomy, Amherst College, 25 East Drive, Amherst, MA 01002, USA}

%% file: FigSet2.tex
\figsetstart
\figsetnum{2}
\figsettitle{NEID activity indicator time series measurements and GLS periodograms.}

\figsetgrpstart
\figsetgrpnum{2.1}
\figsetgrptitle{HD\,4614\,A activity}
\figsetplot{HD4614A_activity_gls.png}
\figsetgrpnote{NEID stellar activity indicators and GLS periodograms}
\figsetgrpend

\figsetgrpstart
\figsetgrpnum{2.2}
\figsetgrptitle{HD\,4628 activity}
\figsetplot{HD4628_activity_gls.png}
\figsetgrpnote{NEID stellar activity indicators and GLS periodograms}
\figsetgrpend

\figsetgrpstart
\figsetgrpnum{2.3}
\figsetgrptitle{HD\,9407 activity}
\figsetplot{HD9407_activity_gls.png}
\figsetgrpnote{NEID stellar activity indicators and GLS periodograms}
\figsetgrpend

\figsetgrpstart
\figsetgrpnum{2.4}
\figsetgrptitle{HD\,10476 activity}
\figsetplot{HD10476_activity_gls.png}
\figsetgrpnote{NEID stellar activity indicators and GLS periodograms}
\figsetgrpend

\figsetgrpstart
\figsetgrpnum{2.5}
\figsetgrptitle{HD\,10700 activity}
\figsetplot{HD10700_activity_gls.png}
\figsetgrpnote{NEID stellar activity indicators and GLS periodograms}
\figsetgrpend

\figsetgrpstart
\figsetgrpnum{2.6}
\figsetgrptitle{HD\,10780 activity}
\figsetplot{HD10780_activity_gls.png}
\figsetgrpnote{NEID stellar activity indicators and GLS periodograms}
\figsetgrpend

\figsetgrpstart
\figsetgrpnum{2.7}
\figsetgrptitle{HD\,19373 activity}
\figsetplot{HD19373_activity_gls.png}
\figsetgrpnote{NEID stellar activity indicators and GLS periodograms}
\figsetgrpend

\figsetgrpstart
\figsetgrpnum{2.8}
\figsetgrptitle{HD\,24496\,A activity}
\figsetplot{HD24496A_activity_gls.png}
\figsetgrpnote{NEID stellar activity indicators and GLS periodograms}
\figsetgrpend

\figsetgrpstart
\figsetgrpnum{2.9}
\figsetgrptitle{HD\,26965 activity}
\figsetplot{HD26965_activity_gls.png}
\figsetgrpnote{NEID stellar activity indicators and GLS periodograms}
\figsetgrpend

\figsetgrpstart
\figsetgrpnum{2.10}
\figsetgrptitle{HD\,34411 activity}
\figsetplot{HD34411_activity_gls.png}
\figsetgrpnote{NEID stellar activity indicators and GLS periodograms}
\figsetgrpend

\figsetgrpstart
\figsetgrpnum{2.11}
\figsetgrptitle{HD\,38858 activity}
\figsetplot{HD38858_activity_gls.png}
\figsetgrpnote{NEID stellar activity indicators and GLS periodograms}
\figsetgrpend

\figsetgrpstart
\figsetgrpnum{2.12}
\figsetgrptitle{HD\,50692 activity}
\figsetplot{HD50692_activity_gls.png}
\figsetgrpnote{NEID stellar activity indicators and GLS periodograms}
\figsetgrpend

\figsetgrpstart
\figsetgrpnum{2.13}
\figsetgrptitle{HD\,51419 activity}
\figsetplot{HD51419_activity_gls.png}
\figsetgrpnote{NEID stellar activity indicators and GLS periodograms}
\figsetgrpend

\figsetgrpstart
\figsetgrpnum{2.14}
\figsetgrptitle{HD\,52711 activity}
\figsetplot{HD52711_activity_gls.png}
\figsetgrpnote{NEID stellar activity indicators and GLS periodograms}
\figsetgrpend

\figsetgrpstart
\figsetgrpnum{2.15}
\figsetgrptitle{HD\,55575 activity}
\figsetplot{HD55575_activity_gls.png}
\figsetgrpnote{NEID stellar activity indicators and GLS periodograms}
\figsetgrpend

\figsetgrpstart
\figsetgrpnum{2.16}
\figsetgrptitle{HD\,68017 activity}
\figsetplot{HD68017_activity_gls.png}
\figsetgrpnote{NEID stellar activity indicators and GLS periodograms}
\figsetgrpend

\figsetgrpstart
\figsetgrpnum{2.17}
\figsetgrptitle{HD\,86728 activity}
\figsetplot{HD86728_activity_gls.png}
\figsetgrpnote{NEID stellar activity indicators and GLS periodograms}
\figsetgrpend

\figsetgrpstart
\figsetgrpnum{2.18}
\figsetgrptitle{HD\,95735 activity}
\figsetplot{HD95735_activity_gls.png}
\figsetgrpnote{NEID stellar activity indicators and GLS periodograms}
\figsetgrpend

\figsetgrpstart
\figsetgrpnum{2.19}
\figsetgrptitle{HD\,110897 activity}
\figsetplot{HD110897_activity_gls.png}
\figsetgrpnote{NEID stellar activity indicators and GLS periodograms}
\figsetgrpend

\figsetgrpstart
\figsetgrpnum{2.20}
\figsetgrptitle{HD\,115617 activity}
\figsetplot{HD115617_activity_gls.png}
\figsetgrpnote{NEID stellar activity indicators and GLS periodograms}
\figsetgrpend

\figsetgrpstart
\figsetgrpnum{2.21}
\figsetgrptitle{HD\,116442 activity}
\figsetplot{HD116442_activity_gls.png}
\figsetgrpnote{NEID stellar activity indicators and GLS periodograms}
\figsetgrpend

\figsetgrpstart
\figsetgrpnum{2.22}
\figsetgrptitle{HD\,126053 activity}
\figsetplot{HD126053_activity_gls.png}
\figsetgrpnote{NEID stellar activity indicators and GLS periodograms}
\figsetgrpend

\figsetgrpstart
\figsetgrpnum{2.23}
\figsetgrptitle{HD\,127334 activity}
\figsetplot{HD127334_activity_gls.png}
\figsetgrpnote{NEID stellar activity indicators and GLS periodograms}
\figsetgrpend

\figsetgrpstart
\figsetgrpnum{2.24}
\figsetgrptitle{HD\,143761 activity}
\figsetplot{HD143761_activity_gls.png}
\figsetgrpnote{NEID stellar activity indicators and GLS periodograms}
\figsetgrpend

\figsetgrpstart
\figsetgrpnum{2.25}
\figsetgrptitle{HD\,146233 activity}
\figsetplot{HD146233_activity_gls.png}
\figsetgrpnote{NEID stellar activity indicators and GLS periodograms}
\figsetgrpend

\figsetgrpstart
\figsetgrpnum{2.26}
\figsetgrptitle{HD\,154345 activity}
\figsetplot{HD154345_activity_gls.png}
\figsetgrpnote{NEID stellar activity indicators and GLS periodograms}
\figsetgrpend

\figsetgrpstart
\figsetgrpnum{2.27}
\figsetgrptitle{HD\,157214 activity}
\figsetplot{HD157214_activity_gls.png}
\figsetgrpnote{NEID stellar activity indicators and GLS periodograms}
\figsetgrpend

\figsetgrpstart
\figsetgrpnum{2.28}
\figsetgrptitle{HD\,166620 activity}
\figsetplot{HD166620_activity_gls.png}
\figsetgrpnote{NEID stellar activity indicators and GLS periodograms}
\figsetgrpend

\figsetgrpstart
\figsetgrpnum{2.29}
\figsetgrptitle{HD\,168009 activity}
\figsetplot{HD168009_activity_gls.png}
\figsetgrpnote{NEID stellar activity indicators and GLS periodograms}
\figsetgrpend

\figsetgrpstart
\figsetgrpnum{2.30}
\figsetgrptitle{HD\,170657 activity}
\figsetplot{HD170657_activity_gls.png}
\figsetgrpnote{NEID stellar activity indicators and GLS periodograms}
\figsetgrpend

\figsetgrpstart
\figsetgrpnum{2.31}
\figsetgrptitle{HD\,172051 activity}
\figsetplot{HD172051_activity_gls.png}
\figsetgrpnote{NEID stellar activity indicators and GLS periodograms}
\figsetgrpend

\figsetgrpstart
\figsetgrpnum{2.32}
\figsetgrptitle{HD\,179957 activity}
\figsetplot{HD179957_activity_gls.png}
\figsetgrpnote{NEID stellar activity indicators and GLS periodograms}
\figsetgrpend

\figsetgrpstart
\figsetgrpnum{2.33}
\figsetgrptitle{HD\,182572 activity}
\figsetplot{HD182572_activity_gls.png}
\figsetgrpnote{NEID stellar activity indicators and GLS periodograms}
\figsetgrpend

\figsetgrpstart
\figsetgrpnum{2.34}
\figsetgrptitle{HD\,185144 activity}
\figsetplot{HD185144_activity_gls.png}
\figsetgrpnote{NEID stellar activity indicators and GLS periodograms}
\figsetgrpend

\figsetgrpstart
\figsetgrpnum{2.35}
\figsetgrptitle{HD\,186408 activity}
\figsetplot{HD186408_activity_gls.png}
\figsetgrpnote{NEID stellar activity indicators and GLS periodograms}
\figsetgrpend

\figsetgrpstart
\figsetgrpnum{2.36}
\figsetgrptitle{HD\,186427 activity}
\figsetplot{HD186427_activity_gls.png}
\figsetgrpnote{NEID stellar activity indicators and GLS periodograms}
\figsetgrpend

\figsetgrpstart
\figsetgrpnum{2.37}
\figsetgrptitle{HD\,187923 activity}
\figsetplot{HD187923_activity_gls.png}
\figsetgrpnote{NEID stellar activity indicators and GLS periodograms}
\figsetgrpend

\figsetgrpstart
\figsetgrpnum{2.38}
\figsetgrptitle{HD\,190360 activity}
\figsetplot{HD190360_activity_gls.png}
\figsetgrpnote{NEID stellar activity indicators and GLS periodograms}
\figsetgrpend

\figsetgrpstart
\figsetgrpnum{2.39}
\figsetgrptitle{HD\,201091 activity}
\figsetplot{HD201091_activity_gls.png}
\figsetgrpnote{NEID stellar activity indicators and GLS periodograms}
\figsetgrpend

\figsetgrpstart
\figsetgrpnum{2.40}
\figsetgrptitle{HD\,217107 activity}
\figsetplot{HD217107_activity_gls.png}
\figsetgrpnote{NEID stellar activity indicators and GLS periodograms}
\figsetgrpend

\figsetgrpstart
\figsetgrpnum{2.41}
\figsetgrptitle{HD\,221354 activity}
\figsetplot{HD221354_activity_gls.png}
\figsetgrpnote{NEID stellar activity indicators and GLS periodograms}
\figsetgrpend

\figsetend

%% file: FigSet8.tex
\figsetstart
\figsetnum{8}
\figsettitle{\texttt{RVSearch} results for stars with no known long-period companions. No trend allowed.}

\figsetgrpstart
\figsetgrpnum{8.1}
\figsetgrptitle{HD\,4628 \texttt{RVSearch} no trend}
\figsetplot{HD4628_rvsearch_notrend.pdf}
\figsetgrpnote{\texttt{RVSearch} results for stars with no known long-period companions. No trend allowed.}
\figsetgrpend

\figsetgrpstart
\figsetgrpnum{8.2}
\figsetgrptitle{HD\,9407 \texttt{RVSearch} no trend}
\figsetplot{HD9407_rvsearch_notrend.pdf}
\figsetgrpnote{\texttt{RVSearch} results for stars with no known long-period companions. No trend allowed.}
\figsetgrpend

\figsetgrpstart
\figsetgrpnum{8.3}
\figsetgrptitle{HD\,10476 \texttt{RVSearch} no trend}
\figsetplot{HD10476_rvsearch_notrend.pdf}
\figsetgrpnote{\texttt{RVSearch} results for stars with no known long-period companions. No trend allowed.}
\figsetgrpend

\figsetgrpstart
\figsetgrpnum{8.4}
\figsetgrptitle{HD\,10700 \texttt{RVSearch} no trend}
\figsetplot{HD10700_rvsearch_notrend.pdf}
\figsetgrpnote{\texttt{RVSearch} results for stars with no known long-period companions. No trend allowed.}
\figsetgrpend

\figsetgrpstart
\figsetgrpnum{8.5}
\figsetgrptitle{HD\,10780 \texttt{RVSearch} no trend}
\figsetplot{HD10780_rvsearch_notrend.pdf}
\figsetgrpnote{\texttt{RVSearch} results for stars with no known long-period companions. No trend allowed.}
\figsetgrpend

\figsetgrpstart
\figsetgrpnum{8.6}
\figsetgrptitle{HD\,19373 \texttt{RVSearch} no trend}
\figsetplot{HD19373_rvsearch_notrend.pdf}
\figsetgrpnote{\texttt{RVSearch} results for stars with no known long-period companions. No trend allowed.}
\figsetgrpend

\figsetgrpstart
\figsetgrpnum{8.7}
\figsetgrptitle{HD\,26965 \texttt{RVSearch} no trend}
\figsetplot{HD26965_rvsearch_notrend.pdf}
\figsetgrpnote{\texttt{RVSearch} results for stars with no known long-period companions. No trend allowed.}
\figsetgrpend

\figsetgrpstart
\figsetgrpnum{8.8}
\figsetgrptitle{HD\,34411 \texttt{RVSearch} no trend}
\figsetplot{HD34411_rvsearch_notrend.pdf}
\figsetgrpnote{\texttt{RVSearch} results for stars with no known long-period companions. No trend allowed.}
\figsetgrpend

\figsetgrpstart
\figsetgrpnum{8.9}
\figsetgrptitle{HD\,38858 \texttt{RVSearch} no trend}
\figsetplot{HD38858_rvsearch_notrend.pdf}
\figsetgrpnote{\texttt{RVSearch} results for stars with no known long-period companions. No trend allowed.}
\figsetgrpend

\figsetgrpstart
\figsetgrpnum{8.10}
\figsetgrptitle{HD\,50692 \texttt{RVSearch} no trend}
\figsetplot{HD50692_rvsearch_notrend.pdf}
\figsetgrpnote{\texttt{RVSearch} results for stars with no known long-period companions. No trend allowed.}
\figsetgrpend

\figsetgrpstart
\figsetgrpnum{8.11}
\figsetgrptitle{HD\,51419 \texttt{RVSearch} no trend}
\figsetplot{HD51419_rvsearch_notrend.pdf}
\figsetgrpnote{\texttt{RVSearch} results for stars with no known long-period companions. No trend allowed.}
\figsetgrpend

\figsetgrpstart
\figsetgrpnum{8.12}
\figsetgrptitle{HD\,52711 \texttt{RVSearch} no trend}
\figsetplot{HD52711_rvsearch_notrend.pdf}
\figsetgrpnote{\texttt{RVSearch} results for stars with no known long-period companions. No trend allowed.}
\figsetgrpend

\figsetgrpstart
\figsetgrpnum{8.13}
\figsetgrptitle{HD\,55575 \texttt{RVSearch} no trend}
\figsetplot{HD55575_rvsearch_notrend.pdf}
\figsetgrpnote{\texttt{RVSearch} results for stars with no known long-period companions. No trend allowed.}
\figsetgrpend

\figsetgrpstart
\figsetgrpnum{8.14}
\figsetgrptitle{HD\,86728 \texttt{RVSearch} no trend}
\figsetplot{HD86728_rvsearch_notrend.pdf}
\figsetgrpnote{\texttt{RVSearch} results for stars with no known long-period companions. No trend allowed.}
\figsetgrpend

\figsetgrpstart
\figsetgrpnum{8.15}
\figsetgrptitle{HD\,110897 \texttt{RVSearch} no trend}
\figsetplot{HD110897_rvsearch_notrend.pdf}
\figsetgrpnote{\texttt{RVSearch} results for stars with no known long-period companions. No trend allowed.}
\figsetgrpend

\figsetgrpstart
\figsetgrpnum{8.16}
\figsetgrptitle{HD\,115617 \texttt{RVSearch} no trend}
\figsetplot{HD115617_rvsearch_notrend.pdf}
\figsetgrpnote{\texttt{RVSearch} results for stars with no known long-period companions. No trend allowed.}
\figsetgrpend

\figsetgrpstart
\figsetgrpnum{8.17}
\figsetgrptitle{HD\,126053 \texttt{RVSearch} no trend}
\figsetplot{HD126053_rvsearch_notrend.pdf}
\figsetgrpnote{\texttt{RVSearch} results for stars with no known long-period companions. No trend allowed.}
\figsetgrpend

\figsetgrpstart
\figsetgrpnum{8.18}
\figsetgrptitle{HD\,127334 \texttt{RVSearch} no trend}
\figsetplot{HD127334_rvsearch_notrend.pdf}
\figsetgrpnote{\texttt{RVSearch} results for stars with no known long-period companions. No trend allowed.}
\figsetgrpend

\figsetgrpstart
\figsetgrpnum{8.19}
\figsetgrptitle{HD\,143761 \texttt{RVSearch} no trend}
\figsetplot{HD143761_rvsearch_notrend.pdf}
\figsetgrpnote{\texttt{RVSearch} results for stars with no known long-period companions. No trend allowed.}
\figsetgrpend

\figsetgrpstart
\figsetgrpnum{8.20}
\figsetgrptitle{HD\,146233 \texttt{RVSearch} no trend}
\figsetplot{HD146233_rvsearch_notrend.pdf}
\figsetgrpnote{\texttt{RVSearch} results for stars with no known long-period companions. No trend allowed.}
\figsetgrpend

\figsetgrpstart
\figsetgrpnum{8.21}
\figsetgrptitle{HD\,157214 \texttt{RVSearch} no trend}
\figsetplot{HD157214_rvsearch_notrend.pdf}
\figsetgrpnote{\texttt{RVSearch} results for stars with no known long-period companions. No trend allowed.}
\figsetgrpend

\figsetgrpstart
\figsetgrpnum{8.22}
\figsetgrptitle{HD\,166620 \texttt{RVSearch} no trend}
\figsetplot{HD166620_rvsearch_notrend.pdf}
\figsetgrpnote{\texttt{RVSearch} results for stars with no known long-period companions. No trend allowed.}
\figsetgrpend

\figsetgrpstart
\figsetgrpnum{8.23}
\figsetgrptitle{HD\,168009 \texttt{RVSearch} no trend}
\figsetplot{HD168009_rvsearch_notrend.pdf}
\figsetgrpnote{\texttt{RVSearch} results for stars with no known long-period companions. No trend allowed.}
\figsetgrpend

\figsetgrpstart
\figsetgrpnum{8.24}
\figsetgrptitle{HD\,170657 \texttt{RVSearch} no trend}
\figsetplot{HD170657_rvsearch_notrend.pdf}
\figsetgrpnote{\texttt{RVSearch} results for stars with no known long-period companions. No trend allowed.}
\figsetgrpend

\figsetgrpstart
\figsetgrpnum{8.25}
\figsetgrptitle{HD\,172051 \texttt{RVSearch} no trend}
\figsetplot{HD172051_rvsearch_notrend.pdf}
\figsetgrpnote{\texttt{RVSearch} results for stars with no known long-period companions. No trend allowed.}
\figsetgrpend

\figsetgrpstart
\figsetgrpnum{8.26}
\figsetgrptitle{HD\,182572 \texttt{RVSearch} no trend}
\figsetplot{HD182572_rvsearch_notrend.pdf}
\figsetgrpnote{\texttt{RVSearch} results for stars with no known long-period companions. No trend allowed.}
\figsetgrpend

\figsetgrpstart
\figsetgrpnum{8.27}
\figsetgrptitle{HD\,185144 \texttt{RVSearch} no trend}
\figsetplot{HD185144_rvsearch_notrend.pdf}
\figsetgrpnote{\texttt{RVSearch} results for stars with no known long-period companions. No trend allowed.}
\figsetgrpend

\figsetgrpstart
\figsetgrpnum{8.28}
\figsetgrptitle{HD\,187923 \texttt{RVSearch} no trend}
\figsetplot{HD187923_rvsearch_notrend.pdf}
\figsetgrpnote{\texttt{RVSearch} results for stars with no known long-period companions. No trend allowed.}
\figsetgrpend

\figsetgrpstart
\figsetgrpnum{8.29}
\figsetgrptitle{HD\,221354 \texttt{RVSearch} no trend}
\figsetplot{HD221354_rvsearch_notrend.pdf}
\figsetgrpnote{\texttt{RVSearch} results for stars with no known long-period companions. No trend allowed.}
\figsetgrpend

\figsetend

%% file: FigSet9.tex
\figsetstart
\figsetnum{9}
\figsettitle{Blind \texttt{RVSearch} results for stars with known long-period companions.}

\figsetgrpstart
\figsetgrpnum{9.1}
\figsetgrptitle{HD\,4614\,A \texttt{RVSearch} blind}
\figsetplot{HD4614A_rvsearch_blind.pdf}
\figsetgrpnote{Blind \texttt{RVSearch} results for stars with known long-period companions.}
\figsetgrpend

\figsetgrpstart
\figsetgrpnum{9.2}
\figsetgrptitle{HD\,24496\,A \texttt{RVSearch} blind}
\figsetplot{HD24496A_rvsearch_blind.pdf}
\figsetgrpnote{Blind \texttt{RVSearch} results for stars with known long-period companions.}
\figsetgrpend

\figsetgrpstart
\figsetgrpnum{9.3}
\figsetgrptitle{HD\,68017 \texttt{RVSearch} blind}
\figsetplot{HD68017_rvsearch_blind.pdf}
\figsetgrpnote{Blind \texttt{RVSearch} results for stars with known long-period companions.}
\figsetgrpend

\figsetgrpstart
\figsetgrpnum{9.4}
\figsetgrptitle{HD\,95735 \texttt{RVSearch} blind}
\figsetplot{HD95735_rvsearch_blind.pdf}
\figsetgrpnote{Blind \texttt{RVSearch} results for stars with known long-period companions.}
\figsetgrpend

\figsetgrpstart
\figsetgrpnum{9.5}
\figsetgrptitle{HD\,116442 \texttt{RVSearch} blind}
\figsetplot{HD116442_rvsearch_blind.pdf}
\figsetgrpnote{Blind \texttt{RVSearch} results for stars with known long-period companions.}
\figsetgrpend

\figsetgrpstart
\figsetgrpnum{9.6}
\figsetgrptitle{HD\,154345 \texttt{RVSearch} blind}
\figsetplot{HD154345_rvsearch_blind.pdf}
\figsetgrpnote{Blind \texttt{RVSearch} results for stars with known long-period companions.}
\figsetgrpend

\figsetgrpstart
\figsetgrpnum{9.7}
\figsetgrptitle{HD\,179957 \texttt{RVSearch} blind}
\figsetplot{HD179957_rvsearch_blind.pdf}
\figsetgrpnote{Blind \texttt{RVSearch} results for stars with known long-period companions.}
\figsetgrpend

\figsetgrpstart
\figsetgrpnum{9.8}
\figsetgrptitle{HD\,186408 \texttt{RVSearch} blind}
\figsetplot{HD186408_rvsearch_blind.pdf}
\figsetgrpnote{Blind \texttt{RVSearch} results for stars with known long-period companions.}
\figsetgrpend

\figsetgrpstart
\figsetgrpnum{9.9}
\figsetgrptitle{HD\,186427 \texttt{RVSearch} blind}
\figsetplot{HD186427_rvsearch_blind.pdf}
\figsetgrpnote{Blind \texttt{RVSearch} results for stars with known long-period companions.}
\figsetgrpend

\figsetgrpstart
\figsetgrpnum{9.10}
\figsetgrptitle{HD\,190360 \texttt{RVSearch} blind}
\figsetplot{HD190360_rvsearch_blind.pdf}
\figsetgrpnote{Blind \texttt{RVSearch} results for stars with known long-period companions.}
\figsetgrpend

\figsetgrpstart
\figsetgrpnum{9.11}
\figsetgrptitle{HD\,201091 \texttt{RVSearch} blind}
\figsetplot{HD201091_rvsearch_blind.pdf}
\figsetgrpnote{Blind \texttt{RVSearch} results for stars with known long-period companions.}
\figsetgrpend

\figsetgrpstart
\figsetgrpnum{9.12}
\figsetgrptitle{HD\,217107 \texttt{RVSearch} blind}
\figsetplot{HD217107_rvsearch_blind.pdf}
\figsetgrpnote{Blind \texttt{RVSearch} results for stars with known long-period companions.}
\figsetgrpend

\figsetend

%% file: FigSet10.tex
\figsetstart
\figsetnum{10}
\figsettitle{Informed \texttt{RVSearch} results for stars with known long-period companions.}

\figsetgrpstart
\figsetgrpnum{10.1}
\figsetgrptitle{HD\,4614\,A \texttt{RVSearch} informed}
\figsetplot{HD4614A_rvsearch_informed.pdf}
\figsetgrpnote{Informed \texttt{RVSearch} results for stars with known long-period companions.}
\figsetgrpend

\figsetgrpstart
\figsetgrpnum{10.2}
\figsetgrptitle{HD\,24496\,A \texttt{RVSearch} informed}
\figsetplot{HD24496A_rvsearch_informed.pdf}
\figsetgrpnote{Informed \texttt{RVSearch} results for stars with known long-period companions.}
\figsetgrpend

\figsetgrpstart
\figsetgrpnum{10.3}
\figsetgrptitle{HD\,68017 \texttt{RVSearch} informed}
\figsetplot{HD68017_rvsearch_informed.pdf}
\figsetgrpnote{Informed \texttt{RVSearch} results for stars with known long-period companions.}
\figsetgrpend

\figsetgrpstart
\figsetgrpnum{10.4}
\figsetgrptitle{HD\,95735 \texttt{RVSearch} informed}
\figsetplot{HD95735_rvsearch_informed.pdf}
\figsetgrpnote{Informed \texttt{RVSearch} results for stars with known long-period companions.}
\figsetgrpend

\figsetgrpstart
\figsetgrpnum{10.5}
\figsetgrptitle{HD\,116442 \texttt{RVSearch} informed}
\figsetplot{HD116442_rvsearch_informed.pdf}
\figsetgrpnote{Informed \texttt{RVSearch} results for stars with known long-period companions.}
\figsetgrpend

\figsetgrpstart
\figsetgrpnum{10.6}
\figsetgrptitle{HD\,154345 \texttt{RVSearch} informed}
\figsetplot{HD154345_rvsearch_informed.pdf}
\figsetgrpnote{Informed \texttt{RVSearch} results for stars with known long-period companions.}
\figsetgrpend

\figsetgrpstart
\figsetgrpnum{10.7}
\figsetgrptitle{HD\,179957 \texttt{RVSearch} informed}
\figsetplot{HD179957_rvsearch_informed.pdf}
\figsetgrpnote{Informed \texttt{RVSearch} results for stars with known long-period companions.}
\figsetgrpend

\figsetgrpstart
\figsetgrpnum{10.8}
\figsetgrptitle{HD\,186408 \texttt{RVSearch} informed}
\figsetplot{HD186408_rvsearch_informed.pdf}
\figsetgrpnote{Informed \texttt{RVSearch} results for stars with known long-period companions.}
\figsetgrpend

\figsetgrpstart
\figsetgrpnum{10.9}
\figsetgrptitle{HD\,186427 \texttt{RVSearch} informed}
\figsetplot{HD186427_rvsearch_informed.pdf}
\figsetgrpnote{Informed \texttt{RVSearch} results for stars with known long-period companions.}
\figsetgrpend

\figsetgrpstart
\figsetgrpnum{10.10}
\figsetgrptitle{HD\,190360 \texttt{RVSearch} informed}
\figsetplot{HD190360_rvsearch_informed.pdf}
\figsetgrpnote{Informed \texttt{RVSearch} results for stars with known long-period companions.}
\figsetgrpend

\figsetgrpstart
\figsetgrpnum{10.11}
\figsetgrptitle{HD\,201091 \texttt{RVSearch} informed}
\figsetplot{HD201091_rvsearch_informed.pdf}
\figsetgrpnote{Informed \texttt{RVSearch} results for stars with known long-period companions.}
\figsetgrpend

\figsetgrpstart
\figsetgrpnum{10.12}
\figsetgrptitle{HD\,217107 \texttt{RVSearch} informed}
\figsetplot{HD217107_rvsearch_informed.pdf}
\figsetgrpnote{Informed \texttt{RVSearch} results for stars with known long-period companions.}
\figsetgrpend

\figsetend

%% file: FigSet11.tex
\figsetstart
\figsetnum{11}
\figsettitle{\texttt{RVSearch} results for stars with no known long-period companions. Trend allowed.}

\figsetgrpstart
\figsetgrpnum{11.1}
\figsetgrptitle{HD\,4628 \texttt{RVSearch} trend}
\figsetplot{HD4628_rvsearch_trend.pdf}
\figsetgrpnote{\texttt{RVSearch} results for stars with no known long-period companions. Trend allowed.}
\figsetgrpend

\figsetgrpstart
\figsetgrpnum{11.2}
\figsetgrptitle{HD\,9407 \texttt{RVSearch} trend}
\figsetplot{HD9407_rvsearch_trend.pdf}
\figsetgrpnote{\texttt{RVSearch} results for stars with no known long-period companions. Trend allowed.}
\figsetgrpend

\figsetgrpstart
\figsetgrpnum{11.3}
\figsetgrptitle{HD\,10476 \texttt{RVSearch} trend}
\figsetplot{HD10476_rvsearch_trend.pdf}
\figsetgrpnote{\texttt{RVSearch} results for stars with no known long-period companions. Trend allowed.}
\figsetgrpend

\figsetgrpstart
\figsetgrpnum{11.4}
\figsetgrptitle{HD\,10700 \texttt{RVSearch} trend}
\figsetplot{HD10700_rvsearch_trend.pdf}
\figsetgrpnote{\texttt{RVSearch} results for stars with no known long-period companions. Trend allowed.}
\figsetgrpend

\figsetgrpstart
\figsetgrpnum{11.5}
\figsetgrptitle{HD\,10780 \texttt{RVSearch} trend}
\figsetplot{HD10780_rvsearch_trend.pdf}
\figsetgrpnote{\texttt{RVSearch} results for stars with no known long-period companions. Trend allowed.}
\figsetgrpend

\figsetgrpstart
\figsetgrpnum{11.6}
\figsetgrptitle{HD\,19373 \texttt{RVSearch} trend}
\figsetplot{HD19373_rvsearch_trend.pdf}
\figsetgrpnote{\texttt{RVSearch} results for stars with no known long-period companions. Trend allowed.}
\figsetgrpend

\figsetgrpstart
\figsetgrpnum{11.7}
\figsetgrptitle{HD\,26965 \texttt{RVSearch} trend}
\figsetplot{HD26965_rvsearch_trend.pdf}
\figsetgrpnote{\texttt{RVSearch} results for stars with no known long-period companions. Trend allowed.}
\figsetgrpend

\figsetgrpstart
\figsetgrpnum{11.8}
\figsetgrptitle{HD\,34411 \texttt{RVSearch} trend}
\figsetplot{HD34411_rvsearch_trend.pdf}
\figsetgrpnote{\texttt{RVSearch} results for stars with no known long-period companions. Trend allowed.}
\figsetgrpend

\figsetgrpstart
\figsetgrpnum{11.9}
\figsetgrptitle{HD\,38858 \texttt{RVSearch} trend}
\figsetplot{HD38858_rvsearch_trend.pdf}
\figsetgrpnote{\texttt{RVSearch} results for stars with no known long-period companions. Trend allowed.}
\figsetgrpend

\figsetgrpstart
\figsetgrpnum{11.10}
\figsetgrptitle{HD\,50692 \texttt{RVSearch} trend}
\figsetplot{HD50692_rvsearch_trend.pdf}
\figsetgrpnote{\texttt{RVSearch} results for stars with no known long-period companions. Trend allowed.}
\figsetgrpend

\figsetgrpstart
\figsetgrpnum{11.11}
\figsetgrptitle{HD\,51419 \texttt{RVSearch} trend}
\figsetplot{HD51419_rvsearch_trend.pdf}
\figsetgrpnote{\texttt{RVSearch} results for stars with no known long-period companions. Trend allowed.}
\figsetgrpend

\figsetgrpstart
\figsetgrpnum{11.12}
\figsetgrptitle{HD\,52711 \texttt{RVSearch} trend}
\figsetplot{HD52711_rvsearch_trend.pdf}
\figsetgrpnote{\texttt{RVSearch} results for stars with no known long-period companions. Trend allowed.}
\figsetgrpend

\figsetgrpstart
\figsetgrpnum{11.13}
\figsetgrptitle{HD\,55575 \texttt{RVSearch} trend}
\figsetplot{HD55575_rvsearch_trend.pdf}
\figsetgrpnote{\texttt{RVSearch} results for stars with no known long-period companions. Trend allowed.}
\figsetgrpend

\figsetgrpstart
\figsetgrpnum{11.14}
\figsetgrptitle{HD\,86728 \texttt{RVSearch} trend}
\figsetplot{HD86728_rvsearch_trend.pdf}
\figsetgrpnote{\texttt{RVSearch} results for stars with no known long-period companions. Trend allowed.}
\figsetgrpend

\figsetgrpstart
\figsetgrpnum{11.15}
\figsetgrptitle{HD\,110897 \texttt{RVSearch} trend}
\figsetplot{HD110897_rvsearch_trend.pdf}
\figsetgrpnote{\texttt{RVSearch} results for stars with no known long-period companions. Trend allowed.}
\figsetgrpend

\figsetgrpstart
\figsetgrpnum{11.16}
\figsetgrptitle{HD\,115617 \texttt{RVSearch} trend}
\figsetplot{HD115617_rvsearch_trend.pdf}
\figsetgrpnote{\texttt{RVSearch} results for stars with no known long-period companions. Trend allowed.}
\figsetgrpend

\figsetgrpstart
\figsetgrpnum{11.17}
\figsetgrptitle{HD\,126053 \texttt{RVSearch} trend}
\figsetplot{HD126053_rvsearch_trend.pdf}
\figsetgrpnote{\texttt{RVSearch} results for stars with no known long-period companions. Trend allowed.}
\figsetgrpend

\figsetgrpstart
\figsetgrpnum{11.18}
\figsetgrptitle{HD\,127334 \texttt{RVSearch} trend}
\figsetplot{HD127334_rvsearch_trend.pdf}
\figsetgrpnote{\texttt{RVSearch} results for stars with no known long-period companions. Trend allowed.}
\figsetgrpend

\figsetgrpstart
\figsetgrpnum{11.19}
\figsetgrptitle{HD\,143761 \texttt{RVSearch} trend}
\figsetplot{HD143761_rvsearch_trend.pdf}
\figsetgrpnote{\texttt{RVSearch} results for stars with no known long-period companions. Trend allowed.}
\figsetgrpend

\figsetgrpstart
\figsetgrpnum{11.20}
\figsetgrptitle{HD\,146233 \texttt{RVSearch} trend}
\figsetplot{HD146233_rvsearch_trend.pdf}
\figsetgrpnote{\texttt{RVSearch} results for stars with no known long-period companions. Trend allowed.}
\figsetgrpend

\figsetgrpstart
\figsetgrpnum{11.21}
\figsetgrptitle{HD\,157214 \texttt{RVSearch} trend}
\figsetplot{HD157214_rvsearch_trend.pdf}
\figsetgrpnote{\texttt{RVSearch} results for stars with no known long-period companions. Trend allowed.}
\figsetgrpend

\figsetgrpstart
\figsetgrpnum{11.22}
\figsetgrptitle{HD\,166620 \texttt{RVSearch} trend}
\figsetplot{HD166620_rvsearch_trend.pdf}
\figsetgrpnote{\texttt{RVSearch} results for stars with no known long-period companions. Trend allowed.}
\figsetgrpend

\figsetgrpstart
\figsetgrpnum{11.23}
\figsetgrptitle{HD\,168009 \texttt{RVSearch} trend}
\figsetplot{HD168009_rvsearch_trend.pdf}
\figsetgrpnote{\texttt{RVSearch} results for stars with no known long-period companions. Trend allowed.}
\figsetgrpend

\figsetgrpstart
\figsetgrpnum{11.24}
\figsetgrptitle{HD\,170657 \texttt{RVSearch} trend}
\figsetplot{HD170657_rvsearch_trend.pdf}
\figsetgrpnote{\texttt{RVSearch} results for stars with no known long-period companions. Trend allowed.}
\figsetgrpend

\figsetgrpstart
\figsetgrpnum{11.25}
\figsetgrptitle{HD\,172051 \texttt{RVSearch} trend}
\figsetplot{HD172051_rvsearch_trend.pdf}
\figsetgrpnote{\texttt{RVSearch} results for stars with no known long-period companions. Trend allowed.}
\figsetgrpend

\figsetgrpstart
\figsetgrpnum{11.26}
\figsetgrptitle{HD\,182572 \texttt{RVSearch} trend}
\figsetplot{HD182572_rvsearch_trend.pdf}
\figsetgrpnote{\texttt{RVSearch} results for stars with no known long-period companions. Trend allowed.}
\figsetgrpend

\figsetgrpstart
\figsetgrpnum{11.27}
\figsetgrptitle{HD\,185144 \texttt{RVSearch} trend}
\figsetplot{HD185144_rvsearch_trend.pdf}
\figsetgrpnote{\texttt{RVSearch} results for stars with no known long-period companions. Trend allowed.}
\figsetgrpend

\figsetgrpstart
\figsetgrpnum{11.28}
\figsetgrptitle{HD\,187923 \texttt{RVSearch} trend}
\figsetplot{HD187923_rvsearch_trend.pdf}
\figsetgrpnote{\texttt{RVSearch} results for stars with no known long-period companions. Trend allowed.}
\figsetgrpend

\figsetgrpstart
\figsetgrpnum{11.29}
\figsetgrptitle{HD\,221354 \texttt{RVSearch} trend}
\figsetplot{HD221354_rvsearch_trend.pdf}
\figsetgrpnote{\texttt{RVSearch} results for stars with no known long-period companions. Trend allowed.}
\figsetgrpend

\figsetend